\documentclass[runningheads,a4paper]{llncs}

\usepackage[T1]{fontenc}
\usepackage{latexsym,bm}
\usepackage{dsfont}
\usepackage{amssymb}
\usepackage{amsmath}

\usepackage[ruled,vlined,linesnumbered,boxed,commentsnumbered]{algorithm2e}

\usepackage{color}
\usepackage[font=bf,skip=\baselineskip]{caption}
\usepackage[labelformat=simple]{subcaption}
\usepackage{graphicx}
\usepackage{epstopdf}
\usepackage{bibentry}
\usepackage{multirow,dcolumn}
\usepackage{umoline}

\usepackage{xspace,afterpage,cancel}
\usepackage[normalem]{ulem}

\usepackage[inline]{enumitem}
\usepackage[inline]{enumitem}
\newenvironment{enumingen}[5]{\begin{enumerate*}[label=#5),itemjoin={{#2 }},itemjoin*={{#2 #3 }},after={#4},#1]}{\end{enumerate*}}
\newenvironment{enumin}[2][]{\begin{enumingen}{#1}{#2}{and}{.}{\arabic*}}{\end{enumingen}}
\newenvironment{enuminNoAnd}[2][]{\begin{enumingen}{#1}{#2}{}{.}{\arabic*}}{\end{enumingen}}
\newenvironment{enuminii}[2][]{\begin{enumingen}{#1}{#2}{and}{.}{\roman*}}{\end{enumingen}}

\usepackage{hyperref}

\newcommand{\Init}{\ensuremath{\mathit{X_0}}\xspace}
\newcommand{\Uns}{\ensuremath{\mathit{X_{us}}}\xspace}

\newcommand{\R}{\ensuremath{\mathbb{R}}\xspace}

\newcommand{\N}{\ensuremath{\mathds{N}}\xspace}

\newcommand{\emphii}[1]{\textbf{#1}} 

\newcommand{\uconsys}{\ensuremath{\langle X, \vec{f}, \Init, \mathcal{I}, \mathcal{U}\rangle}\xspace}

\newcommand{\SOS}{{SOS}\xspace}
\newcommand{\sys}{\ensuremath{\mathcal{M}}\xspace}

\newcommand{\vect}[1]{\ensuremath{\vec{#1}}\xspace}
\newcommand{\vx}{\ensuremath{\vect{x}}\xspace}

\newcommand{\defeq}{\stackrel{\text{def}}{=}\xspace}
\newcommand{\Lf}{\ensuremath{\mathcal{L}_{\vect{f}}}\xspace}

\begin{document}

\title{Piecewise Robust Barrier Tubes for \\ Nonlinear Hybrid Systems with Uncertainty}

\author{Hui Kong\inst{1}, Ezio Bartocci\inst{3},  Yu Jiang\inst{4},  Thomas A. Henzinger\inst{2}}
\institute{
$^1$  Max-Planck-Institute for Software Systems, Kaiserslautern, Germany\\
$^2$  IST Austria, Klosterneuburg, Austria\\
$^3$ TU Wien, Vienna, Austria \\
$^4$ Tsinghua University, Beijing, China\\
}

\maketitle

\begin{abstract}
Piecewise Barrier Tubes (PBT) is a new technique for flowpipe overapproximation 
for nonlinear systems with polynomial dynamics, which leverages a combination of 
barrier certificates. PBT has advantages over traditional time-step based methods 
in dealing with those nonlinear dynamical systems in which there is a large difference 
in speed between trajectories, producing an overapproximation that is time 
independent.
However, the existing approach for PBT is not efficient due to 
the application of interval methods for enclosure-box computation, and it can only deal 
with continuous dynamical systems without uncertainty.  In this paper, we extend the 
approach with the ability to handle both continuous and hybrid dynamical systems 
with uncertainty that can reside in parameters and/or noise.  
We also improve the efficiency of the method significantly, by avoiding 
the use of interval-based methods for the enclosure-box computation without loosing soundness. 
We have developed a C++ prototype implementing the proposed approach and 
we evaluate it on several benchmarks. The experiments  show that 
our approach is more efficient and precise than other methods in the literature.

\end{abstract}

\let\thefootnote\relax\footnotetext{
This research was supported in part by the Austrian Science Fund (FWF)
under grants S11402-N23, S11405-N23 (RiSE/SHiNE), ADynNet (P28182), and Z211-N23 (Wittgenstein
Award) and the Deutsche Forschungsgemeinschaft project 389792660-TRR 248.}


\section{Introduction}
\label{sec:intro}
Hybrid systems (HS)~\cite{henzinger1996theory} are a suitable 
mathematical framework to model dynamical systems with   
both discrete and continuous dynamics.  This formalism has been 
successfully adopted to design cyber-physical 
systems (CPS) whose behavior is characterized by an embedded 
software monitoring and/or controlling a physical substratum.
Formal verification of HS has indeed a
practical impact in engineering by assuring 
important safety-critical requirements at design-time.  

Despite the great effort to advance the state-of-the-art,  
reachability analysis of HS remains one of the most challenging 
verification tasks.  
Although the problem of reachability analysis is in general 
undecidable~\cite{henzinger1996theory} for HS, 
in the last decade several efficient and scalable semidecidable 
approaches have been proposed to analyse  HS with linear 
dynamics~\cite{frehse2011spaceex,GirardLG08,RayGDBBG15,SchuppAMK17,Gurung2018}.

HS with nonlinear ordinary differential equations (ODEs) 
remains still very challenging to solve because these ODEs do not 
have a closed form solution in general. One common strategy to tackle this problem is to compute an over-approximation (also called flowpipe) that 
contains all the possible trajectories originating from an initial set of 
states within a bounded-time horizon~\cite{chen2013flow,Althoff2016a,duggirala2015c2e2,CyrankaIBJSG17,Cyranka18}.  If the overapproximation does not intersect with the unsafe set of states, then 
the system is safe.  
However, if the overapproximation is too coarse, it may intersect  
the unsafe set of states only due to the approximation errors and then the 
verdict about safety may be inconclusive.
Thus, one of the main problem to address is \emph{how to efficiently compute 
	tight over-approximations of the reachable set of states for nonlinear 
	continuous and hybrid systems}. 

To overcome this problem, in a recent paper~\cite{KongBH18}, we have introduced the notion of Piecewise Barrier Tubes (PBT),  a new flowpipe overapproximation for nonlinear systems with polynomial dynamics. The main idea of this approach is that for each segment of a flowpipe, 
it constructs a coarse box that is big enough to contain the segment and then 
it computes  in the box a set of barrier functions~\cite{prajna2004safety,kong2013exponential} 
which work together to form a tube surrounding the flowpipe. 

PBT has advantages over traditional time-step 
based methods in dealing with those nonlinear dynamical 
systems in which there is a large difference in speed between 
trajectories, producing a tight over-approximation that is time 
independent.  However, the approach in~\cite{KongBH18} 
cannot handle uncertainty and hybrid systems.  In addition,  
the use of interval method for enclosure-box computation 
reduces its efficiency.

In this paper, we extend the approach with the ability to handle both 
continuous and hybrid dynamical systems with uncertainty which can 
reside in parameters and/or noise.  We improve the efficiency of the 
method significantly, by avoiding the use of interval method 
for enclosure-box computation without loosing soundness.  We have developed a C++ prototype 
implementing the proposed approach and we evaluate it on several 
benchmarks.  The experiments show that our approach is more 
efficient and precise than other methods proposed in the literature.

The other existing techniques used to compute a bounded flowpipe are mainly
based on interval method~\cite{nedialkov2006interval} or Taylor model~\cite{BerzM98}.  
Interval method is quite efficient even for 
high dimensional systems~\cite{nedialkov2006interval}, but it
suffers from the \emph{wrapping effect} that arises due 
to an uncontrollable growth of the interval enclosure that accumulates 
overapproximation errors.  The use of Taylor model is more precise because it 
uses a vector of polynomials plus a vector of small intervals to 
symbolically represent the flowpipe.  However, checking the intersection 
with the unsafe region requires generally the use of interval 
method that brings back the wrapping effect. 
In particular, the wrapping effect can explode easily when the flowpipe segment 
over a time interval is stretched drastically due to a large difference 
in speed between individual trajectories.

Only recently,  tools such as 
CLRT~\cite{CyrankaIBJSG17,Cyranka18}, Flow*~\cite{chen2013flow}, 
MathSAT SMT solver~\cite{CimattiGIRS18,CimattiGIRS18SAT},  HySAT/iSAT~\cite{FranzleHTRS07}, dReach~\cite{kong2015dreach},
C2E2~\cite{duggirala2015c2e2} and CORA~\cite{Althoff2016a}, 
have made some progresses in verifying nonlinear continuous and 
hybrid models.  Some of these tools ~\cite{CimattiGIRS18SAT,kong2015dreach,FranzleHTRS07}
are based on decision procedures that overcome the theoretical limits 
in nonlinear theories over the reals.  The main idea is to 
encode the reachability problem for nonlinear systems as 
first-order logic formulas over the real numbers. 
A satisfiability modulo theories (SMT) solver implementing such 
procedures can return either a verdict of unsatisfiability when the unsafe region is not reached
or an inconclusive verdict~\cite{kong2015dreach,FranzleHTRS07} 
such as $\delta$-sat if the problem is satisfiable given a certain precision $\delta$ 
(the same problem may result unsatisfiable by increasing the precision).  
However, in the case of unsatisfiability these tools generally do not 
provide a reachable set representation that explains the verdict. Other techniques for reachability analysis of nonlinear systems include invariant generation~\cite{matringe2010generating,sogokon2016method,KongInvriantclusterhscc2017,sankaranarayanan2010automatic,sankaranarayanan2004constructing}, abstraction and hybridization~\cite{roohi2016hybridization,krilavicius2006hybrid,asarin2007hybridization,prabhakar2016hybridization,GrosuBFGGSB11,Bogomolov2015a}. 

The paper is organized as follows.  Section~\ref{sec:preliminaries} 
presents the necessary preliminaries.  Section~\ref{sec:barrier_handelman} 
shows how to compute robust barrier certificates using linear programming, 
while in Section~\ref{sec:PBTcomputation} we present our  
approach to address the reachability analysis problem of nonlinear continuous 
and hybrid systems with uncertainty.  
Section~\ref{sec:evaluation} provides our experimental results and 
we conclude in Section~\ref{sec:conclusion}. 


\section{Preliminaries}\label{sec:preliminaries}
In this section, we recall some concepts used throughout the paper.  
We first clarify some notation conventions.  If not specified otherwise, 
we use boldface lower case letters to denote vectors, we use $\R$ for 
the real numbers field and $\N$ for the set of natural numbers, and we 
consider multivariate polynomials in $\R[\vec{x}]$, where the components 
of $\vec{x}$ act as indeterminates. In addition, for all the polynomials 
$B(\vec{c},\vx)$, we denote by $\vec{c}$ the vector composed of all the 
$c_i$ and denote by $\vx$ the vector composed of all the remaining 
variables $x_i$ that occur in the polynomial. We use $\R_{\geq 0}$ 
and $\R_{>0}$ to denote the domain of nonnegative real number 
and positive real number respectively.  With an abuse of notation, 
we sometimes use $B(\vx) = 0$ for the semialgebraic set it defines. $\partial S$ denotes the boundary of compact set $S$.

%
%

Next, we present the notation of the Lie derivative, which is widely used 
in the discipline of differential geometry. Let $\vec{f}: \mathbb{R}^n \to
\mathbb{R}^n$ be a continuous vector field such that $\dot{x}_i = f_i(\vec{x})$ 
where $\dot{x}_i$ is the time derivative of $x_i(t)$.

\begin{definition}[Lie derivative]\label{def_Lie}
 For a given polynomial $p\in \R[\vect{x}]$ over $\vect{x}=(x_1,\dots,x_n)$ and a continuous system $\dot{\vect{x}} = \vect{f}$, where $\vect{f}=(f_1,\dots,f_n)$, the \emphii{Lie derivative} of $p\in \R[\vect{x}]$ along $\vec{f}$ is defined as $\Lf p =  \sum_{i=1}^n \frac{\partial{p}}{\partial{x_i}} \cdot f_i $.
\end{definition}

Essentially, the Lie derivative of $p$ is the time derivative of $p$, i.e., reflects the change of $p$ over time.

In this paper, we focus on semialgebraic systems with uncertainty, which is described by the following ODE.

\begin{equation}\label{udiff}
 \dot{\vec{x}}  = \vec{f}(\vec{x}(t), \vec{u}(t))
\end{equation}
where $\vec{f}$ is a vector of polynomial functions, $\vec{x}(t)$ is a solution of the system,  $\vec{u}(t)$ is the vector of uncertain parameters and/or perturbation and $\vec{u}(t)$ is Lipschitz continuous. Note that we do not make a distinction between uncertain parameters and perturbation since we deal with them uniformly. Formally, semialgebraic system with uncertainty is defined as follows.

\begin{definition}[Semialgebraic system with uncertainty]
A \emphii{semialgebraic system with uncertainty} is a $5$-tuple $\sys \stackrel{\mbox{def}}{=} \uconsys$, where
\begin{enumerate}
  \item $X \subseteq \mathbb{R}^n$ is the state space of the system $\sys$,
  \item $\vect{f}\in \R[\vx,\vec{u}]^n$ is locally Lipschitz continuous vector function defining the vector flow as in ODE~\eqref{udiff},
  \item $\Init\subseteq X$ is the initial set, which is semialgebraic~\cite{stengle1974nullstellensatz},
  \item $\mathcal{I}$ is the invariant or domain of the system,
  \item $\mathcal{U}$ is a domain for the uncertain parameters and perturbation, i.e., $\vec{u}(t) \in \mathcal{U}$ 
\end{enumerate}
\end{definition}

The local Lipschitz continuity guarantees the existence and uniqueness of the 
differential equation $\dot{\vec{x}}=\vect{f}$ locally. A trajectory of a semialgebraic system with uncertainty is defined as follows.

\begin{definition}[Trajectory]
Given a semialgebraic system with uncertainty $\sys$, a \emphii{trajectory} originating from a point $\vect{x}_0\in \Init$ to time $T>0$ is a continuous and differentiable function $\vect{\zeta}(\vec{x}_0,t):[0, T)\to \mathbb{R}^n$ such that
  \begin{enumin}{,}
   \item $\vect{\zeta}(x_0,0)=\vect{x}_0$
   \item $\exists \vec{u}(\cdot): \forall \tau \in [0,T)$: $\frac{d\vect{\zeta}}{dt}\big|_{t=\tau} = \vect{f}(\vect{\zeta}(\vec{x}_0,\tau), \vec{u}(\tau))$, where $\vec{u}(\cdot): [0,T)\rightarrow \mathcal{U}$
  \end{enumin}
$T$ is assumed to be within the maximal interval of existence of the solution from $\vect{x}_0$.
\end{definition}

For ease of readability, we also use $\zeta(t)$ for $\zeta(\vec{x}_0,t)$ if it is clear from the context. 


\begin{definition}[Safety]
Given an unsafe set $\Uns \subseteq X$, a semialgebraic system with uncertainty $\sys$ is said to be \emphii{safe} if no trajectory $\vect{\zeta}(\vec{x}_0,t)$ of $\sys$ satisfies that $\exists \tau\in \mathbb{R}_{\geq 0}:\vect{\zeta}(\vec{x}_0,\tau)\in \Uns$, where $\vec{x}_0\in \Init$.
\end{definition}


\section{Robust Barrier Certificate by Linear Programming}\label{sec:barrier_handelman}

A barrier certificate for a continuous dynamics system is a real-valued function $B(\vec{x})$ such that
\begin{enuminNoAnd}{}
	\item the initial set and the unsafe set are located on different sides of the hyper-surface
	 $\mathcal{H} = \{\vec{x}\in\R^n \mid B(\vec{x}) = 0\}$ respectively, and
	\item no trajectory originating from the same side of $\mathcal{H}$ as the initial set can
	 cross through $\mathcal{H}$ to reach the other side
\end{enuminNoAnd} Therefore, the existence of such a function $B(\vec{x})$ can guarantee the safety of the system. The above condition can be formalized using an infinite sequence of 
higher order Lie derivatives~\cite{taly2009deductive}.  Unfortunately, this formalization cannot 
be applied directly to barrier certificate computation. Therefore, a couple of sufficient conditions 
for the above condition have been proposed~\cite{kong2013exponential,liu2011computing,gulwani2008constraint}. 
Most recently, based on the sufficient condition in~\cite{prajna2004safety}, a new approach was proposed to 
overapproximate the flowpipe of nonlinear continuous dynamical systems using combination of 
barrier certificates~\cite{KongBH18}. However, the approach is limited to continuous dynamical 
systems without uncertainty.  To tackle this problem, we extend the approach to deal with 
continuous and hybrid systems with uncertainty. Similarly, we adopt the same barrier certificate 
condition as~\cite{KongBH18}, but we introduce the uncertainty in the barrier certificate condition. 
Note that in order to distinguish it from barrier certificate for dynamical system without uncertainty, 
we call a barrier certificate satisfying the following condition \emph{robust barrier certificate}.
\begin{theorem}\label{thrmrobustbarrier}
	Given an uncertain semialgebraic system $\sys = \uconsys$, let \Uns be the unsafe set, the system is guaranteed to be safe if there exists a real-valued function $B(\vec{x})$ such that
	\begin{align}
	&\forall \vec{x}\in \Init: B(\vec{x}) > 0 \label{eq1} \\
	&\forall (\vec{x},\vec{u}) \in \mathcal{I} \times \mathcal{U}: \mathcal{L}_{\vec{f}} B > 0 \label{eq2} \\
	&\forall \vec{x}\in \Uns: B(\vec{x}) < 0 \label{eq3}
	\end{align}
\end{theorem}


The most common approach to barrier certificate computation is by \SOS programming~\cite{prajna2004safety,kong2013exponential}. 
The idea of this kind of approach is to first relax the original constraints like~\eqref{eq1}--\eqref{eq3} into 
a set of positive semidefinite (PSD) polynomials by applying Putinar representation~\cite{putinar1993positive}, 
which is further relaxed by requiring every PSD polynomial has a sum-of-squares decomposition, which 
can be solved by SOS programming in polynomial time.  However, constructing automatically a set of consistent templates for the barrier certificate as well as the auxiliary 
polynomials is not trivial. In addition, \SOS programming method can yield fake solution sometimes due 
to numerical error.  

An alternative to \SOS programming based approaches is to use linear programming based approaches. 
This class of approaches relies on an LP-relaxation to the original constraint. In~\cite{sankaranarayanan2013lyapunov,ben2015linear}, 
to compute Lyapunov function, an LP-relaxation was obtained by applying Handelman representation to 
the original constraint. Recently, this kind of LP-relaxation was adopted in~\cite{KongBH18} to compute 
piecewise barrier tubes.  In~\cite{yang2016linear}, an extended version of Handelman representation, 
called Krivine representation~\cite{lasserre2005polynomial}, was employed for barrier certificate computation. 
Compared to Handelman representation, which can only deal with convex polytopes, Krivine representation 
can deal with more general compact semialgebraic sets. However, Krivine representation requires 
normalizing the polynomials involved, which is expensive.

In this paper, we adopt the same representation as in~\cite{KongBH18}, i.e., Handelman representation as our LP-relaxation scheme 
for Theorem~\ref{thrmrobustbarrier}. We assume that the initial set $\Init$, the unsafe 
set $\Uns$, the invariant $\mathcal{I}$, the parameter and/or perturbance space are 
all convex and compact polyhedra, i.e., $\Init = \{\vec{x}\in\R^n \mid p_1(\vec{x})\geq 0,\cdots,p_{m_1}(\vec{x})\geq 0\}$, $\mathcal{I} = \{\vec{x}\in\R^n \mid q_1(\vec{x})\geq 0,\cdots,q_{m_2}(\vec{x})\geq 0\}$, $\mathcal{U} = \{\vec{u}\in\R^l \mid w_1(\vec{u})\geq 0,\cdots,w_{m_3}(\vec{u})\geq 0\}$ and $\Uns = \{\vec{x}\in\R^n \mid r_1(\vec{x})\geq 0,\cdots,r_{m_4}(\vec{x})\geq 0\}$
where $p_i(\vec{x})$, $q_i(\vec{x})$, $r_k(\vec{x})$ and $w_i(\vec{u})$, are all linear polynomials. Then, Theorem~\ref{thrmrobustbarrier} can be relaxed as follows.
\begin{theorem}\label{thrmhandeltrans}
Given a semialgebraic system with uncertainty $\sys = \uconsys$, let $\Init$, $\Uns$, $\mathcal{I}$ and $\mathcal{U}$ be defined as above, the system is guaranteed to be safe if there exists a real-valued polynomial function $B(\vec{x})$ such that 
\begin{align}
B(\vec{x}) &\equiv \sum_{|\vec{\alpha}|\leq M_1} \lambda_{\vec{\alpha}} \prod_{i=1}^{m_1}p_i^{\alpha_i} + \epsilon_1\label{eq4}\\
\mathcal{L}_{\vec{f}} B  &\equiv   \sum_{|\vec{\beta}|\leq M_2} \lambda_{\vec{\beta}}\prod_{i=1}^{m_2}q_i^{\beta_i}\prod_{j=1}^{m_3}w_j^{\beta_{m_2+j}} + \epsilon_2  \label{eq5}\\
-B(\vec{x}) &\equiv \sum_{|\vec{\gamma}|\leq M_3} \lambda_{\vec{\gamma}}\prod_{i=1}^{m_4}r_i^{\gamma_i} + \epsilon_3 \label{eq6}
\end{align}
where $\vec{\alpha}=(\alpha_k), \vec{\beta}=(\beta_k), \vec{\gamma} = (\gamma_k)$,  $\lambda_{\vec{\alpha}}, \lambda_{\vec{\beta}}, \lambda_{\vec{\gamma}}\in \R_{\geq 0}$, $\epsilon_i \in \R_{>0}$ and $M_i\in\N, i=1,\cdots,3$.
\end{theorem}


\begin{remark}
Theorem~\ref{thrmhandeltrans} implies that the system $\sys$ can be proved to be safe as long as we can find a real-valued polynomial function $B(\vec{x})$ such that $B(x)$, $-B(\vec{x})$ and  $\mathcal{L}_{\vec{f}} B$ can be written as a nonnegative combination of the products of the powers of the polynomials defining $\Init$, $\Uns$ and $\mathcal{I}\times \mathcal{U}$ respectively. This theorem provides us with a solution to solve barrier certificate by linear programming. Given a polynomial template $B(\vec{c},\vec{x})$ for $B(\vec{x})$, where $\vec{c}$ is the coefficients of the monomials  to be decided in $B(\vec{c},\vec{x})$, we substitute $B(\vec{c},\vec{x})$ for $B(\vec{x})$ occurring in the conditions~\eqref{eq4}--\eqref{eq6} to obtain three polynomial identities in $\R[\vec{x}]$ with linear polynomials in $\R[\vec{c},\vec{\lambda}]$ as their coefficients, where $\vec{\lambda}$ is a vector composed of all the $\lambda_{\vec{\alpha}}, \lambda_{\vec{\beta}}, \lambda_{\vec{\gamma}}$ occurring in \eqref{eq4}--\eqref{eq6}. Since \eqref{eq4}--\eqref{eq6} are identities, then all the coefficients of the corresponding monomials on both sides of the identities must be identical. By collecting the corresponding coefficients of the monomials on both sides of the identities and let them equal respectively, we obtain a system $S$ of linear equations and inequalities on $\vec{c},\vec{\lambda}$. Now, finding a robust barrier certificate is converted to finding a feasible solution for $S$, which can be solved by linear programming efficiently. Since the degree of $B(\vec{c},\vec{x})$ is key to the expressive power of $B(\vec{c},\vec{x})$, in our implementation, we attempt to solve a barrier certificate from a group of templates with different degrees.
\end{remark}

Due to the page limit, we do not elaborate on our algorithm 
for barrier certificate computation, but we demonstrate how it 
works in the following example.


\begin{example}\label{exam1}
Given a 2D system defined by $\dot{x} = 2x + 3y + u_1, \dot{y} = -4x + 2y + u_2$, let $\Init=\{(x,y)\in\R^2\mid p_1 = x + 100 \geq 0, p_2= -90 - x \geq 0, p_3 = y + 45 \geq 0, p_4 = -40 - y 
\geq 0\}$, $\mathcal{I} =\{(x,y)\in\R^2\mid q_1 = x + 110 \geq 0, q_2 = -80 - x \geq 0, q_3 = y + 45 \geq 0, q_4 = -20 - y 
\geq 0\}$, $\mathcal{U} = \{(u_1,u_2)\in\R^2 \mid w_1 = u_1 + 50.0 \geq 0, w_2 = 50.0 - u_1\geq 0, w_3 = u_2 + 50.0 \geq 0, w_4 = 50.0 - u_2 \geq 0\}$ and $\Uns=\{(x,y)\in\R^2\mid r_1 = x + 98 \geq 0, r_2 = -90 - x \geq 0, r_3 = y + 24 \geq 0, r_4 = -20 - y 
\geq 0\}$. Assume $B(\vec{c},\vec{x}) = c_0 + c_1x + c_2y$, $M_i = \epsilon_i = 1$ for $i=1,\cdots,3$,  then we obtain the following polynomial identities according to Theorem~\ref{thrmhandeltrans}
\begin{align*}
&c_1 + c_2x + c_3y - \sum_{i=1}^4 \lambda_{1i} p_i -\epsilon_1 \equiv 0\\
&c_2 (2x + 3y + u_1) + c_3(-4x + 2y + u_2) - \sum_{j=1}^4 \lambda_{2j} q_j -  \sum_{j=1}^4 \lambda_{3j} w_j -\epsilon_2 \equiv 0\\
&-(c_1 + c_2x + c_3y) - \sum_{k=1}^4 \lambda_{4k} r_k -\epsilon_3 \equiv 0
\end{align*}
where $\lambda_{ij}\geq 0$ for $i,j=1,\cdots,4$.
If we collect the coefficients of $x,y, u_1, u_2$ in the above polynomials and let them be $0$, we obtain a system $S$ of linear polynomial 
equations and inequalities over $c_i, \lambda_{ij}$. By solving $S$ using linear programming, we obtain a feasible 
solution with $c_1 = -1263.5, c_2 = -11.5, c_3 = -5.85$.


\end{example}
%
%


\section{Piecewise Robust Barrier Tubes}\label{sec:PBTcomputation}
The idea of piecewise robust barrier tubes (PRBTs) is to use robust barrier tubes (RBTs) to piecewise overapproximate the flowpipe segments of nonlinear hybrid systems with uncertainty, where each RBT is essentially a cluster 
of robust barrier certificates which are situated around the flowpipe to form a tight tube enclosing the flowpipe. The basic idea of PRBT computation is shown in Algorithm~\ref{algo4PRBT}.

\begin{algorithm}[!ht]
	\SetKwData{PRBT}{PRBT}
	\SetKwData{RBT}{RBT}
	\SetKwData{E}{E}
	\SetKwData{N}{N}
	\SetKwData{Found}{Found}
	\SetKwData{FAILED}{FAILED}
	\SetKwFunction{Push}{Push}
	\SetKwFunction{Length}{Length}
	\SetKwInOut{Input}{input}
	\SetKwInOut{Output}{output}
	\caption{PRBT computation}\label{algo4PRBT}
	\Input{$\vec{f}$: dynamics of the system; \Init: Initial set;
		$\mathcal{U}$: set of uncertainty; 
		$\N$: number of robust barrier tubes (\RBT) in \PRBT; $(\theta_{min},d_{min})$: parameters for simulation
	}
	\Output{\PRBT: piecewise robust barrier tube}
	\BlankLine
	$\PRBT \leftarrow$ empty queue\; 
	\While{$\Length(\PRBT) < \N$}
	{
			$[\Found, \theta, d] \leftarrow [false, \theta_0, d_0]$ \;
		\While{$\theta > \theta_{min}$}
		{
			$\E \leftarrow$ construct a coarse enclsoure-box for $\Init$ by $(\theta,d)$-simulation\; \label{enccomp}
			$[\Found, \RBT,\Init'] \leftarrow $compute \RBT inside $E$ and obtain a set $\Init'  \supseteq (\RBT \cap \partial \E)$ \;
			\eIf{not \Found}
			{
				$(\theta,d) \leftarrow 1/2*(\theta,d)$;	 \tcp{to shrink \E}
				continue;
			}
			{
				$\PRBT  \leftarrow \Push(\PRBT, \RBT)$; \tcp{add RBT to the queue of PRBT}
				$\Init \leftarrow \Init' $ ; \tcp{update \Init for computing next RBT}
				break;
			}	
		}
			  \lIf{not \Found}{break  }
	}
	return \PRBT\;
\end{algorithm}

\subsection{Construction of the Enclosure-box}
A key step in PRBT computation is the construction of enclosure-box for a given compact initial 
set. Note that here \emph{an enclosure-box is a hyperrectangle that entirely contains a flowpipe segment}. In principle, the smaller the enclosure-box, 
the easier it is to compute a barrier tube.  However, to make full use of the power of 
nonlinear overapproximation, it is desirable to have as big enclosure-box as possible 
so that fewer barrier tubes are needed to cover a flowpipe.

In~\cite{KongBH18},  interval method was adopted to build an enclosure-box.  However, the main problem with interval method 
is that the enclosure-box thus computed is usually very small which will result in a big number of barrier tubes for a fixed length of flowpipe. On the one hand, this will lead to an increasing burden on barrier tube computation. On the other hand, the capability of barrier tube in overapproximating complex flowpipe can not be fully released.  For these reasons, we choose 
to use a purely simulation-based approach without losing soundness.
 
A key concept involved in our simulation-based enclosure-box construction is 
\emph{twisting of trajectory}, which is a measure of maximal bending of trajectories in a box. 
For the convenience of presentation, we present the formal definition of {twisting of trajectory} 
as follows. 

\begin{definition}[Twisting of trajectory]\label{def_trajectory_twist}
	Let \sys be a continuous system and $\zeta(t)$ be a trajectory of \sys.
	Then, $\zeta(t)$ is said to have a twisting of $\theta$ on the time interval $I = [T_1, T_2]$, written as $\xi_I(\zeta)$,
	if it satisfies that $\xi_I(\zeta) = \theta$, where $
	\xi_I(\zeta) \defeq \sup_{t_1,t_2 \in I} \arccos\bigg(\frac{\langle \dot{\zeta}(t_1), \dot{\zeta}(t_2) \rangle}{\|\dot{\zeta}(t_1)\|\|\dot{\zeta}(t_2)\|}\bigg)$.    	
\end{definition}

Then, we have Algorithm~\ref{algo4box} to compute enclosure-box.

\begin{algorithm}[!ht]
	\SetKwData{G}{G}
	\SetKwData{E}{E}
	\SetKwData{P}{P}
	\SetKwData{succ}{succ}
	\SetKwData{counter}{counter}
	\SetKwFunction{Simul}{Simul}
	\SetKwFunction{Dist}{Dist}
	\SetKwFunction{NormalForm}{NormalForm}
	\SetKwInOut{Input}{input}
	\SetKwInOut{Output}{output}
	\caption{Construct enclosure-box}\label{algo4box}
	\Input{$\vec{f}(\vec{x},\vec{u})$: system dynamics; \Init: initial set; $\mathcal{U}$: uncertain parameters;  $\theta$: twisting of simulation; $\theta_{min}$: minimal theta for simulation; $d$: maximum distance of simulation;}
	\Output{\E: an enclosure-box containing $\Init$; \P: plane where flowpipe exits \;
		 \G: range of intersection of $Flow_f(\Init)$ with plane $P$ by simulation}
	\BlankLine
	$\vec{u}_c \leftarrow $ center point of $\mathcal{U}$\; \label{lbl_ctrpt} 
	$\vec{f}_c(\vec{x}) \leftarrow$ the center dynamic $\vec{f}(\vec{x},\vec{u}_c)$\; \label{lbl_ctrdyn} 
	$S_0 \leftarrow$ sample a set of points from $\Init$\; \label{lbl_samplepts}
	select a point $\vec{x}_0 \in S_0$\; \label{lbl_simctrpt}
	$\succ \leftarrow false$\; 
	\While{$\theta \geq \theta_{min}$ \label{splitloop}}
	{
		$\vec{x}_e \leftarrow $ end point of $(\theta,d)$-simulation of $\vec{f}_c(\vec{x}) $ for $\vec{x}_0$ \; \label{lbl_blk_pln}  
		\ForEach{$\vec{x}_e^i$: plane in the $i$'th dimension of $\vec{x}_e$}
		{
			do simulation for all the points in $S_0$, update \G and \E \;
			\If{all the simulations hit $\vec{x}_e^i$}
			{
				 $\P \leftarrow \vec{x}_e^i$\;
				 $\succ \leftarrow true$\;
			}			
		}
		
		\eIf{\succ}
		{
			bloat \E s.t $Flow_{\vec{f}}(\Init)$ exits from \E only through the facet in \P\; \label{lbl_bloat}
			return $[\E,\P,\G]$\; \label{lbl_ret}
		}
		{
			$[\theta,d] \leftarrow 1/2*[\theta,d]$\; \label{lbl_shrink}
				
		}
	}

\end{algorithm}

\begin{remark}
	 In this paper, we assume that both $\Init$ and $\mathcal{U}$ are defined by hyperrectangles. The basic idea of enclosure-box construction is that, given a continuous dynamical system with uncertainty, we first remove the uncertainty by taking the center point $\vec{u}_c$ of $\mathcal{U}$ for the dynamics (line~\ref{lbl_ctrpt}--\ref{lbl_ctrdyn}). Then, we sample a set $S_0$ of points from $\Init$ for simulation (line~\ref{lbl_samplepts}). Prior to doing simulation for $S_0$, we first select a point $\vec{x}_0$ (usually the center point of $\Init$) to do $(\theta,d)$-simulation to obtain the end point $\vec{x}_e$ of the simulation (line~\ref{lbl_blk_pln}). \emph{A $(\theta,d)$-simulation is a simulation that stops either when 
	 the twisting of the simulation reaches $\theta$ or when the Euclidean distance between $\vec{x}_0$ and $\vec{x}_e$ reaches $d$}. The motivation to get the end point $\vec{x}_e$ is that, there are $n$ planes of the form $x_i = \vec{x}_e^i$ (the i'th element of $\vec{x}_e$) intersecting at $\vec{x}_e$, so we want to check if one of the $n$ planes, say $P$, was hit by all the simulations that start from $S_0$, and if yes, it is very likely that $P$ cut through the entire flowpipe. Then, we take $P$ as one of the facets of the desired enclosure-box $\mathtt{E}$. In addition, during the simulations, we simultaneously keep updating
	 \begin{enumin}{,}
	 	\item the boundary where the simulations can reach and use that range as our candidate enclosure-box $\mathtt{E}$
	 	\item the boundary range $\mathtt{G}$ where the simulations intersect with the plane $P: x_i = \vec{x}_e^i$ 
	 \end{enumin} 
	If we end up finding such a plane $P$, we will push the other facets of $\mathtt{E}$ outwards to make the flowpipe exit only from this specific facet of $\mathtt{E}$. Of course, this objective cannot be guaranteed only by simulation and pushing, we need to further check if the flowpipe does not intersect the other facets of $\mathtt{E}$, which can be done according to Theorem~\ref{thrm_enclo}.
\end{remark}

\begin{theorem}\label{thrm_enclo}
Given an uncertain semialgebraic system \sys = \uconsys, assume $E \subset \mathcal{I}$ is an enclosure-box of $\Init$ and $F_i$ is a facet of $E$. The flowpipe of \sys from $\Init$ does not intersect $F_i$, i.e, $(Flow_f(\Init) \cap F_i) \cap E = \emptyset$ if there exists a barrier certificate $B_i(\vec{x})$ for $F_i$ inside $E$.
\end{theorem}

\begin{remark}
Theorem~\ref{thrm_enclo} can be easily proved by the definition of barrier certificate, which is ignored here. In order to make sure that the flowpipe evades a facet $F_i$ of $\mathtt{E}$, according to Theorem~\ref{thrmrobustbarrier}, we only need to find a barrier certificate for $F_i$. In the case of no barrier certificate being found, further bloating to the facet of $E$ will be performed. If bloating facet still end up with failure, we keep shrinking $E$ by setting $(\theta,d)$ to $(\theta/2,d/2)$ until barrier certificates are found for all the facet of $E$ or $\theta$ gets less than some threshold $\theta_{min}$. 
\end{remark}

\subsection{Computation of Robust Barrier Tube}
An ideal application scenario of barrier certificate is when we can prove the 
safety property using a single barrier certificate.  Unfortunately, this is usually 
not true because the flowpipe can be very complicated so that no polynomial 
function of a specified degree satisfies the constraint. In the previous subsection, 
we introduce how to obtain for an initial set $\Init$ an enclosure-box $E$ in 
which the system dynamics is simple enough so that a robust barrier certificate 
$B(\vec{x})$ can be easily computed. Therefore, we can compute a set of 
robust barrier certificates, which 
we call Robust Barrier Tube (RBT), to create a tight overapproximation for 
the flowpipe provided that there is a set of auxiliary sets serving as unsafe 
sets. Formally, we define RBT as follows.

\begin{definition}[Robust Barrier Tube ({RBT})]\label{defbarrtube}
Given a semialgebraic system \sys = \uconsys, let $E$ be an enclosure-box 
of $\Init$ and $X_{AS} = \{X_{AS}^i: X_{AS}^i \subseteq E\}$ be a set of 
auxiliary sets (AS), an {RBT} is a set of real-valued functions 
$\Phi = \{B_i(\vec{x}), i=1,\cdots,m\}$ such that for all $B_i(\vec{x})\in \Phi$:
\begin{enuminii}{,}
 \item $\forall \vec{x}\in \Init: B_i(\vec{x}) > 0$
 \item $\forall (\vec{x},\vec{u})\in E\times \mathcal{U}: \mathcal{L}_{\vec{f}} B_i > 0$
 \item $\forall \vec{x} \in X_{AS}^i: B_i(\vec{x}) < 0$
\end{enuminii}

\end{definition} 

The precision of {RBT} depends closely on the set $X_{AS}$ of ASs.  Therefore, 
to derive a good barrier tube, we need to first construct a set of high quality ASs. 
The factors that could affect the quality of the set $X_{AS}$ of ASs include 
\begin{enumin}{,}
\item the number of {AS}s
\item the position, size and shape of AS
\end{enumin}
Roughly speaking, the more ASs we have, if positioned properly, the more 
precise the RBT would be. Regarding the position, size and shape of AS, 
a desirable AS should 
\begin{enumin}{,}
	\item be as close to the flowpipe as possible
	\item spread widely around the flowpipe 
	\item be shaped like a shell for the flowpipe
\end{enumin}
Intuitively, a high quality set of ASs could be shaped like a ring around a 
human finger so that the barrier tube is tightly confined in the narrow space between the ring and the finger.  With the key factors aforementioned in mind, we developed 
Algorithm~\ref{algo4BT} for RBT computation.

\begin{algorithm}[!ht]
	\SetKwData{RBT}{RBT}
	\SetKwData{E}{E}
	\SetKwData{G}{G}
	\SetKwData{D}{D}
	\SetKwData{P}{P}
	\SetKwData{AS}{AS}
	\SetKwFunction{ComputeRBC}{ComputeRBC}
	\SetKwFunction{Expand}{Expand}
	\SetKwFunction{Push}{Push}
	\SetKwFunction{CreateAS}{CreateAS}
	\SetKwFunction{Contract}{Contract}
	\SetKwFunction{Diff}{Diff}
	\SetKwInOut{Input}{input}
	\SetKwInOut{Output}{output}
	\caption{Compute robust barrier tube}\label{algo4BT}
	\Input{$\vec{f}$: system dynamics; \Init: Initial set;
		  \E: enclosure-box of \Init; 
		  $\mathcal{U}$: set of uncertainty; 
		  \P: plane where flowpipe exits from enclosure-box \E;
		  \G: box approx. of ($\P \cap Flow_{\vec{f}}(\Init)$) by simulation;
		  $\epsilon$: difference between \AS's (auxiliary set)
	      }
	\Output{\RBT: barrier tube; $X_0'$: box over-approx. of ($\RBT \cap E$)}
	\BlankLine

	\ForEach{$G_{ij}$: a facet of \G \label{lbl_main_loop}}
	{
		\AS $\longleftarrow \CreateAS(\G,\P,G_{ij})$\; \label{lbl_crt_as}
		\While{true \label{lbl_start_while}}
		{
			$[found,B_{ij}] \longleftarrow \ComputeRBC(\vec{f},\Init,\E,\AS,\mathcal{U})$\; \label{lbl_compute}
				
			\lIf{$found$}
			{ 
				\AS' $ \longleftarrow$ \Expand(\AS) \label{lbl_expand}
			}
			\lElse
			{
				\AS' $\longleftarrow$ \Contract(\AS) \label{lbl_contract}
			}	
				
			\If{$\Diff(\AS',\AS) \leq \epsilon$}{break\;}
			
			\AS $\longleftarrow$ \AS'\;
		} \label{lbl_end_while}
		  	  		
		\eIf{$found$}
		{
			$\RBT \longleftarrow \Push(\RBT,B_{ij})$\;	 
			break;
		}
		{return FAIL}
		
	}
    return SUCCEED\;
\end{algorithm}

\begin{figure}[t!]
	\begin{subfigure}[b]{0.23\linewidth}
		\centering
		\includegraphics[scale = 0.25 ,trim=5cm 4cm 3cm 9cm,clip]{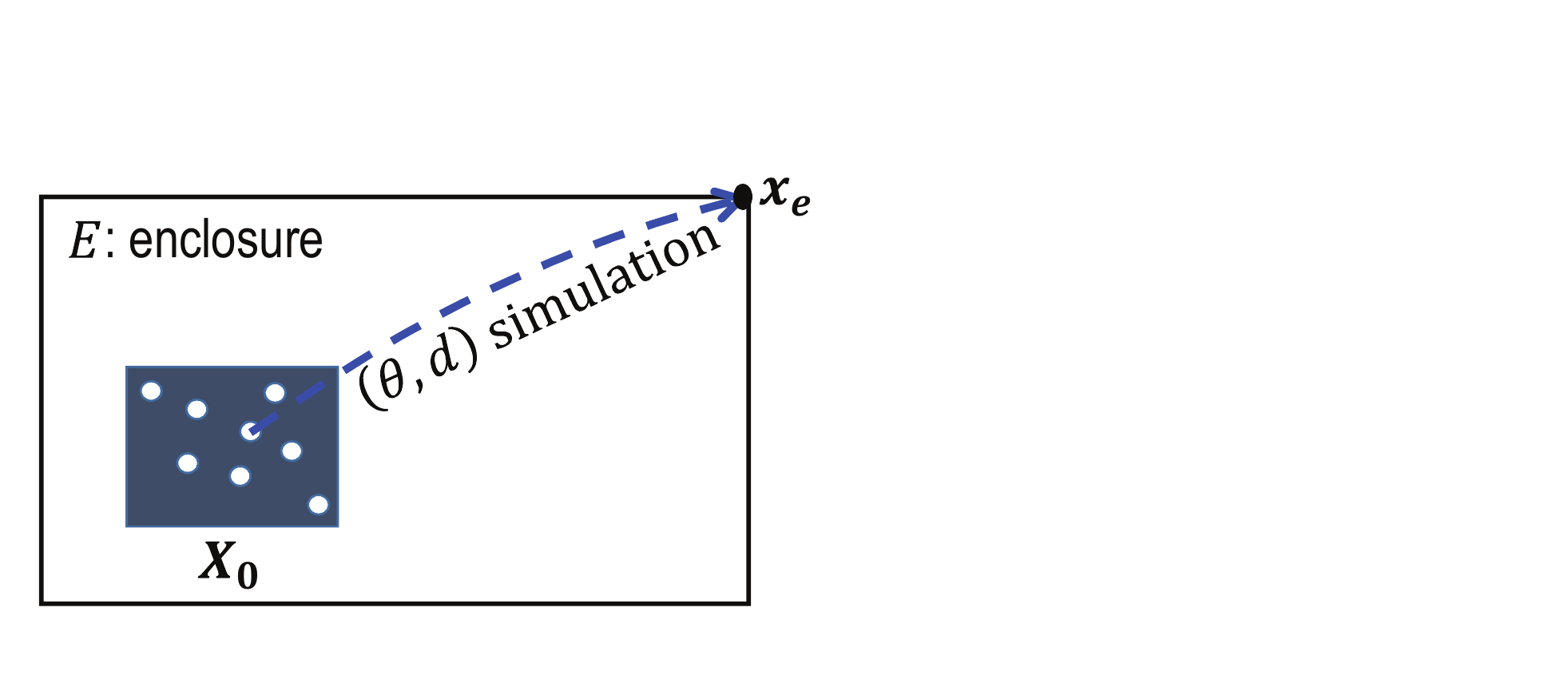}
		\caption{}
		\label{subfig_barr1}
	\end{subfigure}
	\begin{subfigure}[b]{0.23\linewidth}
		\centering
		\includegraphics[scale = 0.25 ,trim=5cm 4cm 3cm 9cm,clip]{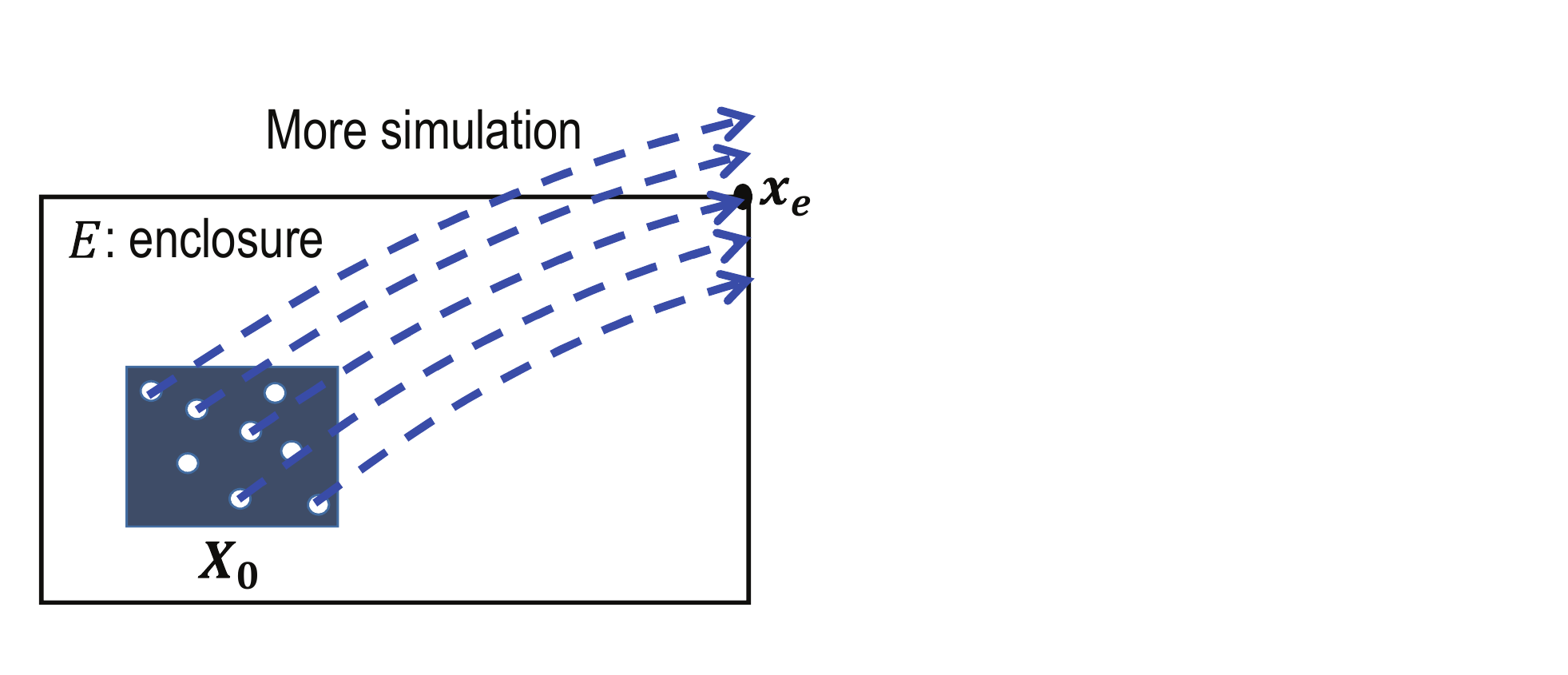}
		\caption{}
		\label{subfig_barr2}
	\end{subfigure}
	\begin{subfigure}[b]{0.23\linewidth}
		\centering
		\includegraphics[scale = 0.25 ,trim=5cm 4cm 3cm 9cm,clip]{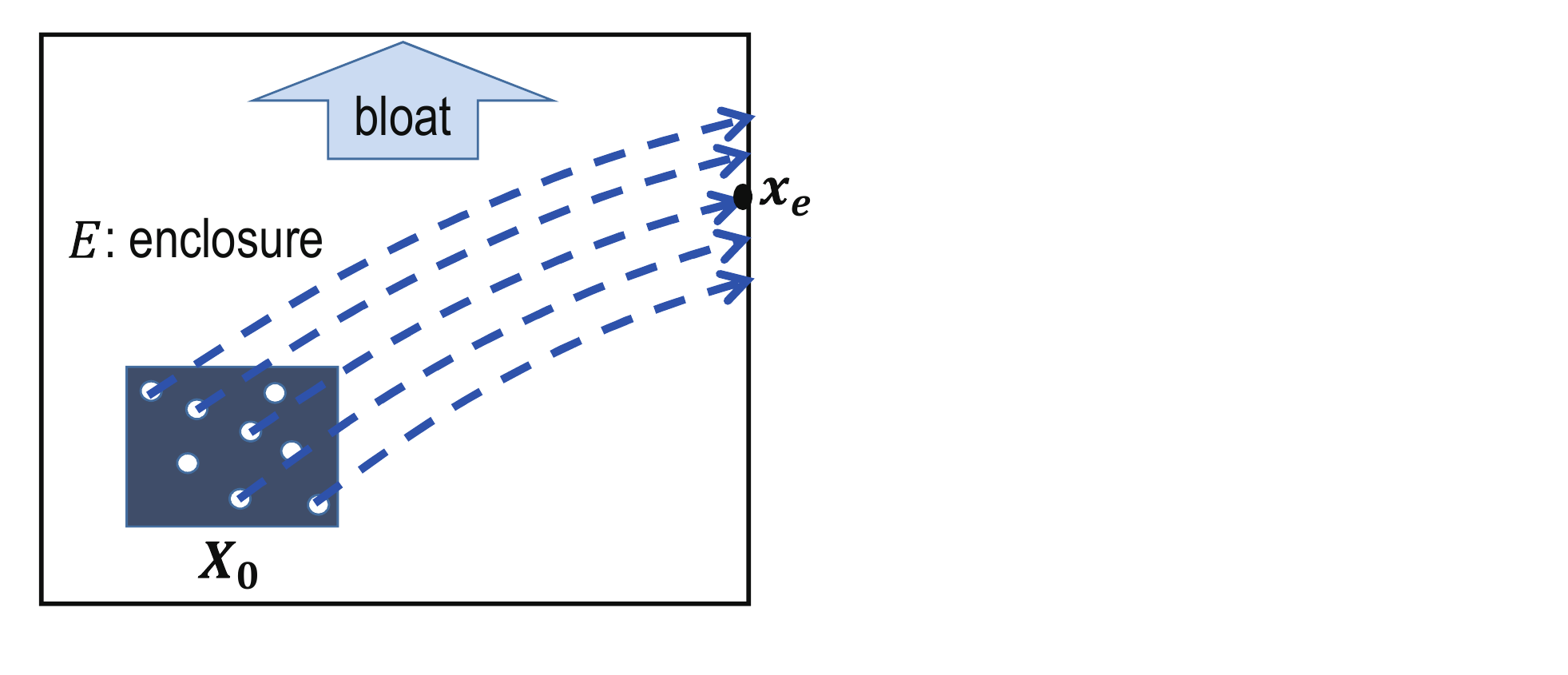}
		\caption{}
		\label{subfig_barr3}
	\end{subfigure}
	\begin{subfigure}[b]{0.23\linewidth}
		\centering
		\includegraphics[scale = 0.25 ,trim=5cm 4cm 3cm 9cm,clip]{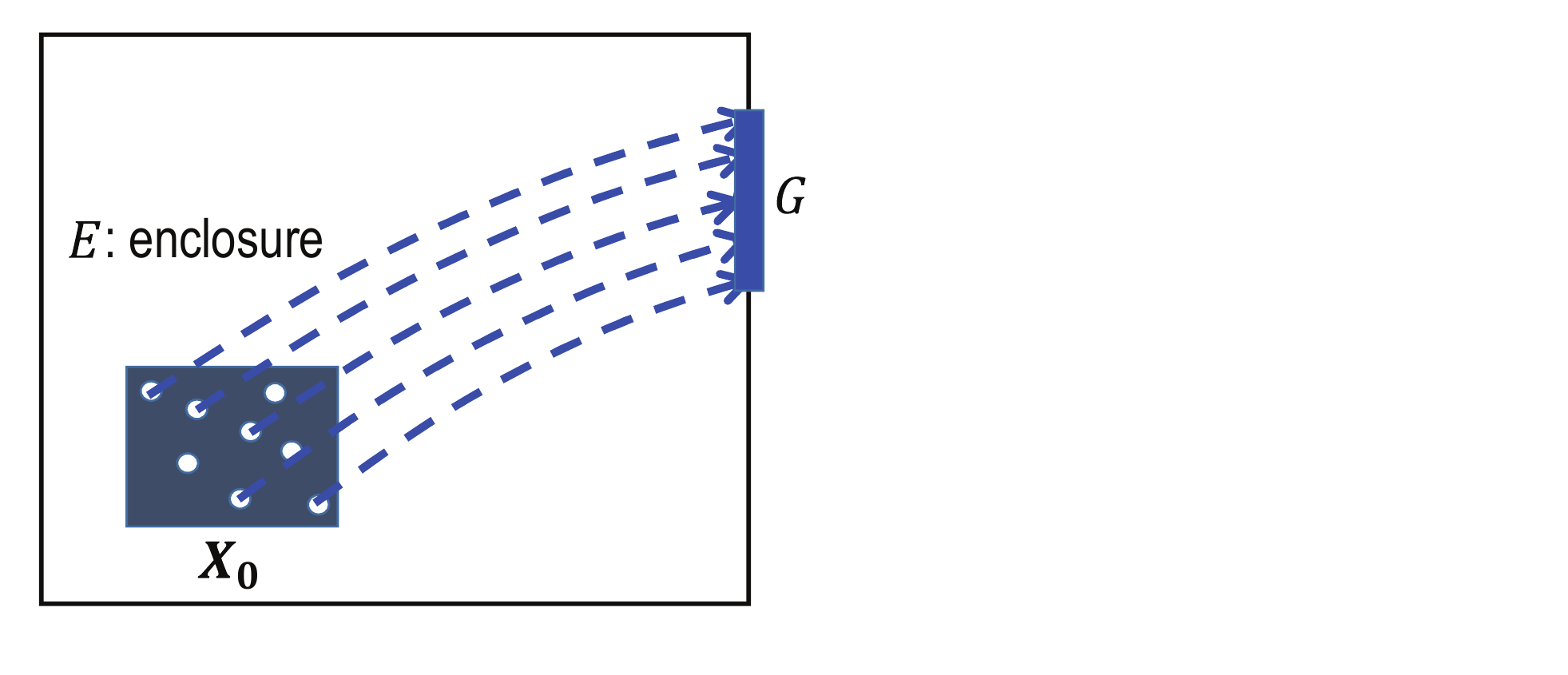}
		\caption{}
		\label{subfig_barr4}
	\end{subfigure}

	\begin{subfigure}[b]{0.23\linewidth}
		\centering
		\includegraphics[scale = 0.25 ,trim=5cm 4cm 3cm 9cm,clip]{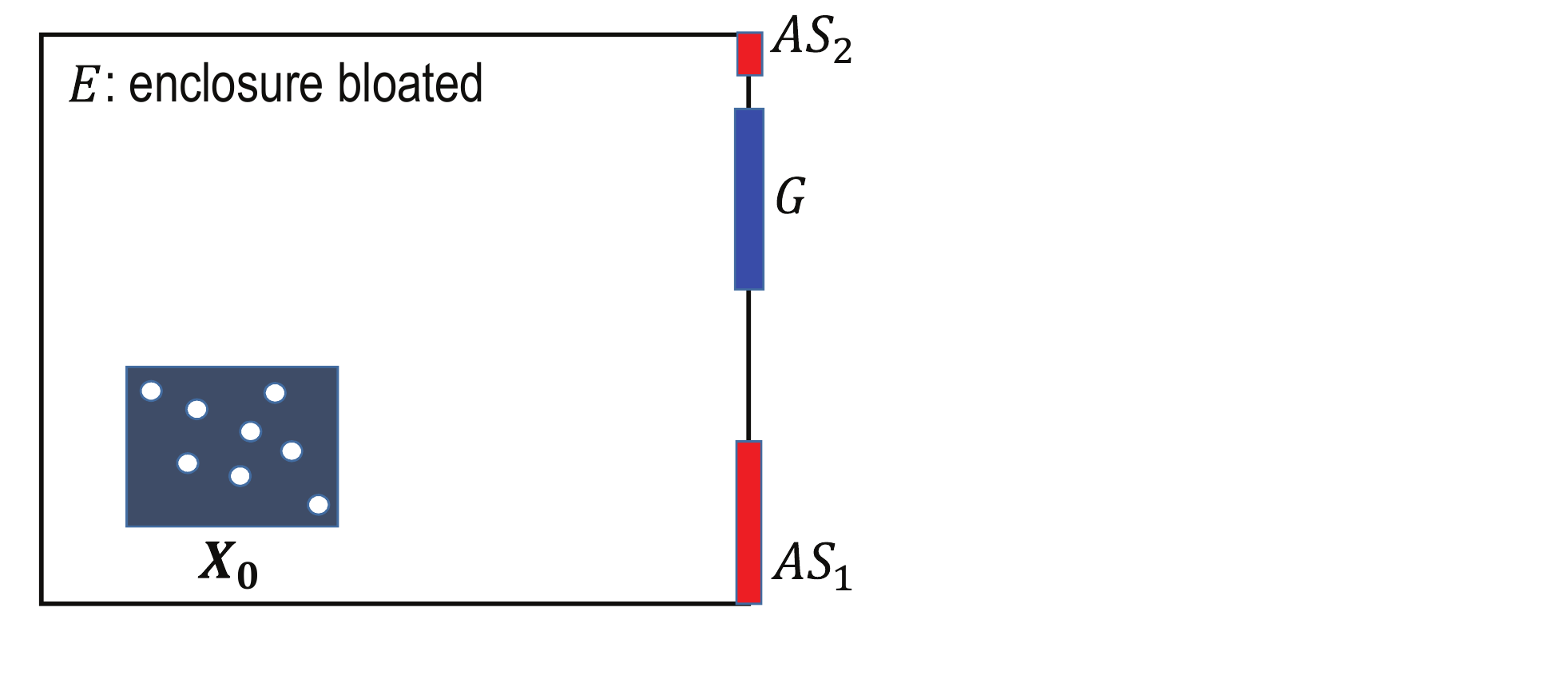}
		\caption{}
		\label{subfig_barr5}
	\end{subfigure}
	\begin{subfigure}[b]{0.43\linewidth}
		\centering
		\includegraphics[scale = 0.25 ,trim=2cm 4cm 3cm 9cm,clip]{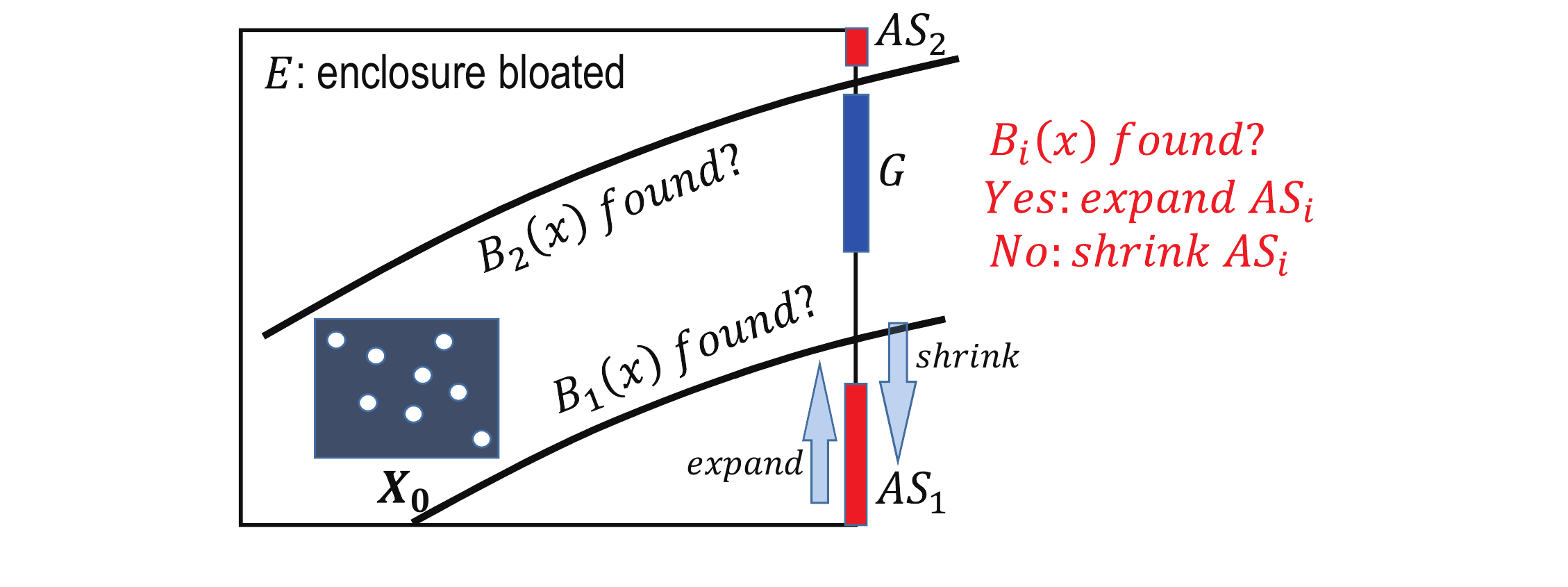}
		\caption{}
		\label{subfig_barr6}
	\end{subfigure}
	\begin{subfigure}[b]{0.23\linewidth}
		\centering
		\includegraphics[scale = 0.25 ,trim=5cm 4cm 3cm 9cm,clip]{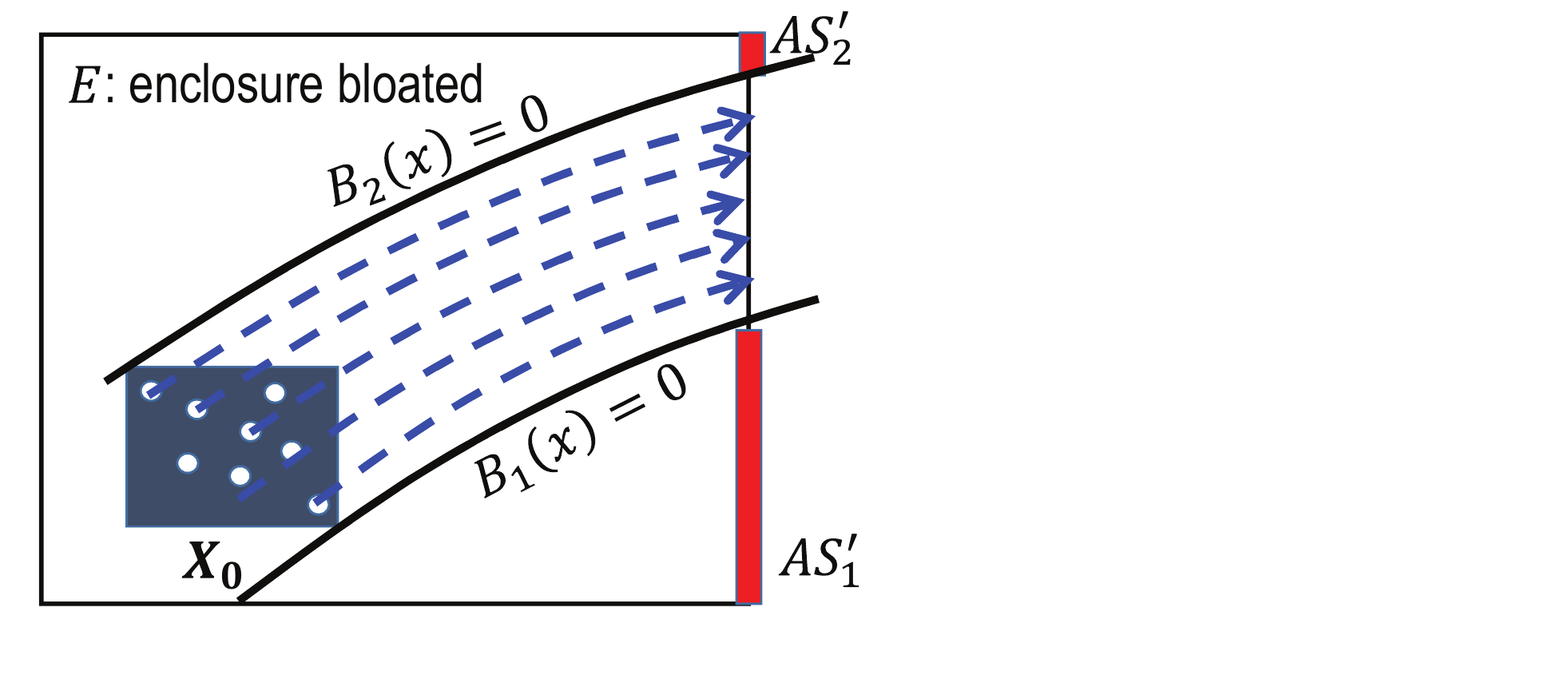}
		\caption{}
		\label{subfig_barr7}
	\end{subfigure}
	\caption{(a)$\rightarrow$(g): demonstration of  {RBT} computation}
	\label{fig:barrtubedemo}
\end{figure}

\begin{remark}
In principle, the more barrier certificates we use, the better overapproximation 
we may achieve.  However, using more barrier certificates also means more 
computation time.  Therefore, we have to make a trade-off between precision 
and efficiency. In Algorithm~\ref{algo4BT}, we choose to use RBT consisting 
of $2(n-1)$ barrier certificates for $n$ dimensional dynamical systems, which 
means we need to construct $2(n-1)$ ASs.  We use the same scheme as 
in~\cite{KongBH18} to construct ASs.  Recall that we get a coarse region 
$\mathtt{G}$ where the flowpipe intersects with one of the facets of $\mathtt{E}$ 
during the construction of the enclosure-box $\mathtt{E}$. Since $\mathtt{G}$ 
is an $n-1$ dimensional box, the RBT must contain $\mathtt{G}$.  Therefore, 
we choose to construct $2(n-1)$ ASs which are able to form a tight hollow 
hyper-rectangle around $\mathtt{G}$. The idea is that for each facet 
$\mathtt{G}_{ij}$ of $\mathtt{G}$, we construct an $n-1$ dimensional 
hyper-rectangle between $\mathtt{G}_{ij}$ and $E_{ij}$ as an $\mathtt{AS}$ (line~\ref{lbl_crt_as}), where $E_{ij}$ is the $n-1$ dimensional face of $\mathtt{E}$ that corresponds to $\mathtt{G}$. 
Then, we use Algorithm~\ref{algo4BT} to compute an RBT(line~\ref{lbl_compute}). 
In the \textbf{while} loop~\ref{lbl_start_while}, we try to find the best barrier 
certificate by adjusting the width of $\mathtt{AS}$ (line~\ref{lbl_expand} 
and~\ref{lbl_contract}) iteratively until the difference in width between two consecutive 
$\mathtt{AS}$s is less than the specified threshold $\epsilon$.  To be intuitive, 
we provide Figure~\ref{fig:barrtubedemo} to demonstrate the process.

\end{remark}

\subsection{PRBT for Continuous Dynamics} 

The idea of computing PRBT is straightforward. Given an initial set $\Init$, we first construct a coarse enclosure-box $\mathtt{E}$ containing $\Init$ and then we further compute an RBT inside $\mathtt{E}$ to get a much more precise overapproximation for the flowpipe. Meanwhile, we obtain a hyper-rectangle $R$ formed by ASs with a hollow $\Init'$ in the middle. Since the intersection of the RBT and the facet of $\mathtt{E}$ is contained entirely in the hollow $\Init'$ of $R$, we use $\Init'$ as a new initial set and repeat the entire process to compute a PRBT step by step. Since our approach is time independent, the length of a PRBT cannot be measured by the length of time horizon. Hence, in our implementation, we try to compute a specified number of RBTs.

\subsection{PRBT for Hybrid Dynamics}
To extend our approach with the ability to deal with hybrid systems, we need to handle two problems
\begin{enuminii}{,}
	\item compute the intersection of RBT and guard set
	\item compute the image of the intersection after discrete jump
\end{enuminii} In general, these two issues can be very hard depending on what kind of guard sets and transitions are defined for the hybrid systems. 

In this paper, we make some assumptions on the hybrid systems under consideration. Let a discrete transition $\tau$ be defined as follows.
\begin{equation}
 \tau_{ll'} = \langle \mathtt{Guard}_{ll'}, \mathtt{Trans}_{ll'} \rangle
\end{equation}
where $l$ and $l'$ are the locations of the dynamics before and after a discrete transition respectively, $\mathtt{Guard}_{ll'} = \{\vec{x}\in \R^n \mid x_i \sim b_i, \sim \in \{\leq, \geq\}\}$ and $\mathtt{Trans}_{ll'}: \vec{x}' = A\vec{x}$, where $A$ is an $n$-dimensional matrix. Based on this assumption, the problem of computing the intersection of RBT and guard set is reduced to computing the intersection of RBT with a plane of $x_i = b_i$, which can be handled using a similar strategy to computing the intersection of RBT with the facet of enclosure-box. Hence, we have Algorithm~\ref{algo4disctrans} to deal with discrete transition of a hybrid system.

\begin{algorithm}[t]
	\SetKwData{BT}{BT}
	\SetKwData{E}{E}
	\SetKwData{G}{G}
	\SetKwData{P}{P}
	\SetKwData{Guard}{Guard}
	\SetKwData{Trans}{Trans}
	\SetKwData{PRBT}{PRBT}
	\SetKwData{InitQ}{InitQ}
	\SetKwData{counter}{counter}
	\SetKwFunction{findRBT}{findRBT}
	\SetKwFunction{ConstructEnc}{ConstructEnc}
	\SetKwFunction{Push}{Push}
	\SetKwFunction{Pop}{Pop}
	\SetKwInOut{Input}{input}
	\SetKwInOut{Output}{output}
	\caption{handle discrete transition of hybrid system}\label{algo4disctrans}
	\Input{$X_0^l$: intermediate initial set at location $l$; 
		$\Guard_{ll'}$: guard set of transition $\tau_{ll'}$;
		 $\Trans_{ll'}$: image mapping of transition $\tau_{ll'}$ }
	\Output{$X_0^{l'}$: image of transition $\tau_{ll'}$}
	\BlankLine
	$\InitQ \leftarrow \Push(\InitQ, X_0^l)$\; 
	\While{$\InitQ$ not empty \label{lbl_intersectloop}}
	{
		$X_0^l \leftarrow \Pop(\InitQ)$\;
		$\E \leftarrow$ construct enclosure-box for $X_0^l$ \;\label{lbl_enclo}
		\If{$\E \cap \Guard_{ll'} == \emptyset$}{continue;}
		$\E \leftarrow \E \cap \overline{\Guard}_{ll'}$\; \label{lbl_cutenc}
		$X_{\Phi\cap \E} \leftarrow $ do simulation and barrier certificate computation to find an overapproximation for the region where the flowpipe $\Phi$ intersects with the guard plane $x_i = b_i$\; \label{lbl_guardintersect}
		$Q_{\Phi\cap\E} \leftarrow \Push(Q_{\Phi\cap\E},X_{\Phi\cap\E}) $\;
		\ForEach{$\E_{ij}$: facet of $\E$ except guard plane}
		{
			$X_0^{ij}\leftarrow $ do simulation and barrier certificate computation to an overapproximation for the region where the barrier tube intersects with $E_{ij}$; \label{lbl_nonguardintersect}
			$\InitQ \leftarrow \Push(\InitQ, X_0^{ij})$	\;		
		}
		
	}
	
	$X_{\Phi\cap \E} \leftarrow $ box overapprox. $Q_{\Phi\cap E}$\;
	$X_0^{l'} \leftarrow \Trans_{ll'}X_{\Phi\cap \E}$ \;
	
\end{algorithm}

\begin{remark}
The strategy to deal with discrete transitions of hybrid systems is that every time we obtain an enclosure-box $\mathtt{E}$, we first detect whether $\mathtt{E}$ intersects with some guard set $\mathtt{Guard_{ll'}}$. If no, we proceed with the normal process of PRBT computation. Otherwise, we switch to the procedure of Algorithm~\ref{algo4disctrans} in which the input $X_0^l$ is the last state set whose enclosure-box intersects with $\mathtt{Guard_{ll'}}$. Since the flowpipe may not cross through the guard plane entirely, we use the {while} loop in line~\ref{lbl_intersectloop} to compute an overapproximation for the intersection. The basic idea of the {while} loop is that, given a state set $X_0^l$, we first construct an enclosure-box $\mathtt{E}$ by simulation(line~\ref{lbl_enclo}), if $\mathtt{E}$ intersects with $\mathtt{Guard_{ll'}}$, we shrink $\mathtt{E}$ by cutting off the part of $\mathtt{E}$ that lies in the guard set (line~\ref{lbl_cutenc}). As a result of this operation, the flowpipe could exit from $\mathtt{E}$ not only through the guard plane but also through other facets of $\mathtt{E}$. For each of those facets, we compute an overapproximation for its intersection with the flowpipe using simulation and barrier certificate computation(line~\ref{lbl_guardintersect} and \ref{lbl_nonguardintersect}). In addition, since those intersections $X_0^{ij}$ that do not lie in the guard plane could still reach the guard plane later, we therefore push them into a queue for further exploration.

\end{remark}
\section{Implementation and Experiments}\label{sec:evaluation}

We have developed PRBT, a software prototype written in C++ 
that implements the concepts and the algorithms presented 
in this paper.  PRBT computes piecewise robust barrier 
tubes for nonlinear continuous and hybrid systems with 
polynomial dynamics.  We compare  our approach in efficiency and precision
 with the state-of-the-art tools Flow* and CORA
using several benchmarks of nonlinear 
continuous and hybrid systems. Note that since C2E2 does not support uncertainty, so we cannot compare with it.
The experiments were carried out on a desktop computer with a 
$3.6$GHz \emph{Intel 8 Core i7-7700} CPU and $32$ GB memory.   

\subsection{Nonlinear Continuous Systems}
We consider six nonlinear benchmark systems with polynomial 
dynamics for which their models and settings are provided 
in Table~\ref{tbl:benchmarks}.
\begin{table*}[h]
	\caption{Continuous dynamical model definitions}
	\centering	
	\begin{tabular}{ | c || l || l || l | }
		\hline
		\multicolumn{1}{|c||}{Model} & \multicolumn{1}{c||}{Dynamics} & \multicolumn{1}{c||}{Uncertainty} & \multicolumn{1}{c|}{\Init} \\
		\hline
		\hline
		Controller 2D             & $\dot{x} = d_1 x y + y^3 + 2$                  & $d_1\in [0.95,1.05]$ &  $x\in [29.9,30.1]$       \tabularnewline
		& $\dot{y} =  d_2 x^2 + 2 x - 3 y\qquad$         & $d_2\in [0.95,1.05]$ &  $y\in [-38,-36]$          \tabularnewline
		\hline
		Van der Pol               & $\dot{x} = y + d_1$                            & $d_1\in [-0.01,0.01]$ & $x\in [1, 1.5] $     \tabularnewline
		Oscillator	              & $\dot{y}  =  y - x - x^2 y + d_2$                & $d_2 \in [-0.01,0.01] $ & $y\in [2.40,2.45]$    \tabularnewline
		\hline
		
		Lotka-Volterra            & $\dot{x} = x (1.5 - y) + d_1$                     & $d_1 \in [-0.01,0.01]$ & $x\in [4.6,5.5]$     \tabularnewline
		& $\dot{y} =  -y (3 - x) - d_2$                     & $d_2 \in [-0.01,0.01]$ & $y\in [1.6,1.7]$     \tabularnewline
		\hline
		Buckling            & $\dot{x} = y + d_1          $                    & $d_1 \in [-0.01,0.01]$ & $x\in [-0.5,-0.4]$     \tabularnewline
		Column  & $\dot{y} = 2 x - x^3 - 0.2 y + 0.1 + d_2$          & $d_1 \in [-0.01,0.01]$ & $y\in [-0.5,-0.4]$     \tabularnewline
		\hline
		Jet Engine                & $\dot{x} = -y - 1.5 x^2 - 0.5 x^3 - 0.5$    & $d_1\in [-0.005, 0.005] $ & $x\in [1.19,1.21]$     \tabularnewline
			                      & $\qquad + d_1$    & $d_2\in [-0.005, 0.005] $ &$y\in [0.8,1.0]$      \tabularnewline
		& $\dot{y} = 3 x - y + d_2$                       &  &     \tabularnewline
		\hline
		
		& $\dot{x} = 10 (y - x) + d_1$         & $d_1 \in [-0.001,0.001]$ &  $x\in [1.79,1.81]$      \tabularnewline
		Controller 3D  			& $\dot{y} = x^3 + d_2$              & $d_2 \in [-0.001,0.001]$& $y\in [1.0,1.1] $       \tabularnewline
		& $\dot{z} = x y - 2.667 z$          & & $z\in [0.5,0.6]$        \tabularnewline
		\hline
	\end{tabular}
	\label{tbl:benchmarks}
\end{table*}

The experimental results are reported in Table~\ref{tbl:data}.  Since our approach is time 
independent, which is different from Flow* and CORA, to make the comparison fair 
enough, we choose to compute a slightly longer flowpipe than the other two tools.  
Note that there are two columns for time for Flow*. The reason why we have an 
extra time column for Flow* is that it can be very fast and precise to compute the Taylor 
model for a given system.  However, Taylor models cannot be in general applied 
directly to solve the safety verification problem.  Checking their intersection with 
the unsafe set requires their transformation  into simpler 
geometric form (e.g. box), which has an exponential complexity in the number of 
the dimensions and it needs to be considered in the overall running time.     

\begin{table*}[!h]
	\caption{Experimental results for benchmark systems. \#var: number of variables; T: computing time for flowpipe; TT: computing time including box transformation; N: 
		number of flowpipe segments; D: candidate degrees for template polynomial (for PRBT only); TH: 
		time horizon for flowpipe (for Flow* and CORA only). F/E: failed to terminate under 30min or exception happened.}
	\centering
	\begin{tabular}{|c||c||c|c|c||c|c|c|c|c|c|c|}
		\hline
		& \multicolumn{1}{c||}{} & \multicolumn{3}{c||}{PRBT} & \multicolumn{1}{c|}{} & \multicolumn{3}{c|}{Flow*} & \multicolumn{2}{c|}{CORA} \\
		\cline{3-5} \cline{7-11}
		Model & \#var &\multicolumn{1}{c|}{T} & \multicolumn{1}{c|}{N} & \multicolumn{1}{c||}{D}& \multicolumn{1}{c|}{TH} & \multicolumn{1}{c|}{T}  & \multicolumn{1}{c|}{TT} & \multicolumn{1}{c|}{N}  & \multicolumn{1}{c|}{T} & \multicolumn{1}{c|}{N}  \tabularnewline
		\hline
		\hline
		Controller 2D    & 2 & 35.73   & 12  & 3   & 0.012  & 48.8  & 417.64     & 240  & F  & 1200       		\tabularnewline
		Van der Pol      & 2 & 221.62  & 17  & 4   & 6.74   & 23.88 & 1111.05    & 135  & E  &   --  			\tabularnewline
		Lotka-Volterra   & 2 & 30.10   & 9   & 4   & 3.20    & 6.06  & 405.32     & 320    &  40.30 &  160         \tabularnewline
		Buckling Column  & 2 & 74.02   & 35  & 3   & 14.00     & F & -- & --  & 734.81  & 1400             \tabularnewline
		Jet Engine       & 2 & 240.98  & 18  & 4   & 9.50    & F & -- & --  &   1.69 &  190            \tabularnewline
		Controller 3D    & 3 & 774.58  & 20  & 3   & 0.55   & F & -- & --  &  F  & --			    \tabularnewline
		\hline
	\end{tabular}
	\label{tbl:data}
\end{table*}

\begin{figure}[t!]
	\begin{subfigure}[b]{0.21\linewidth}
		\centering
		\includegraphics[height=18mm,keepaspectratio]{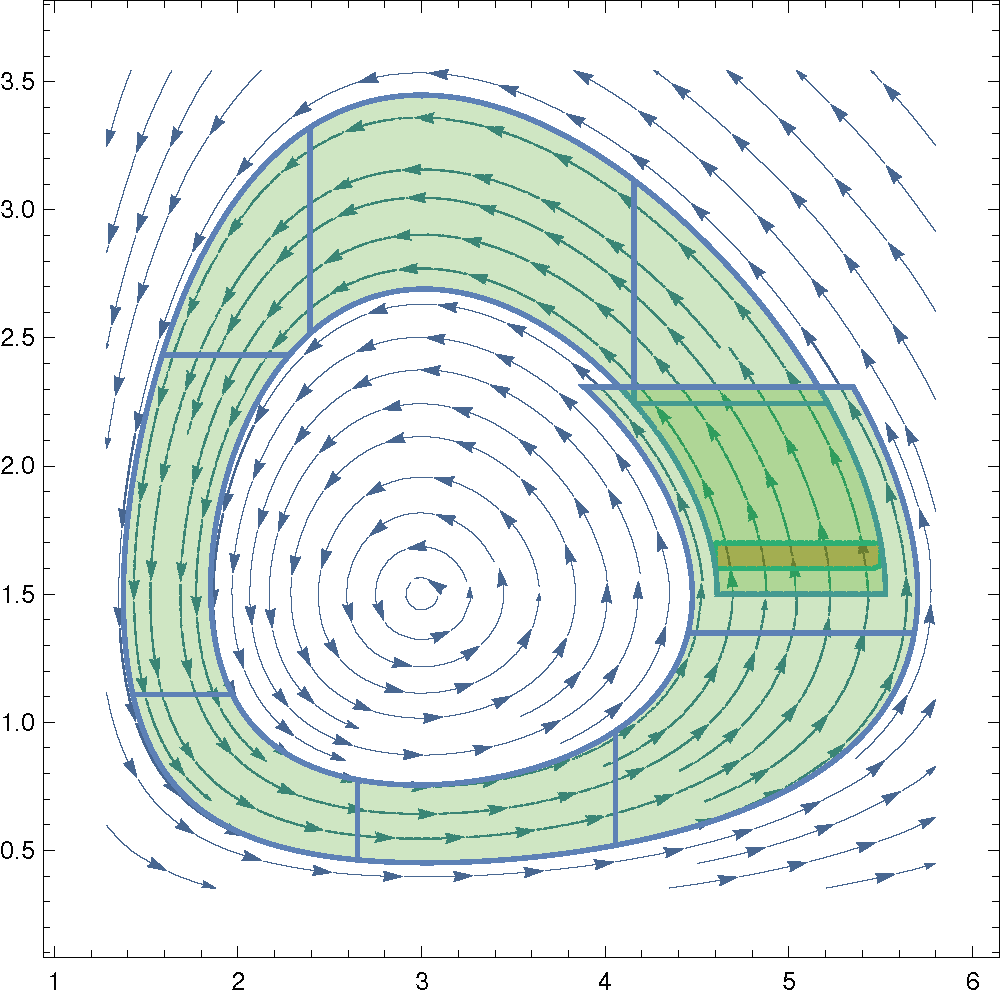}
		\caption{PRBT}
		\label{Lotka-Vloterra-PBTS}
	\end{subfigure}
	\begin{subfigure}[b]{0.21\linewidth}
		\centering
		\includegraphics[height=18mm,keepaspectratio]{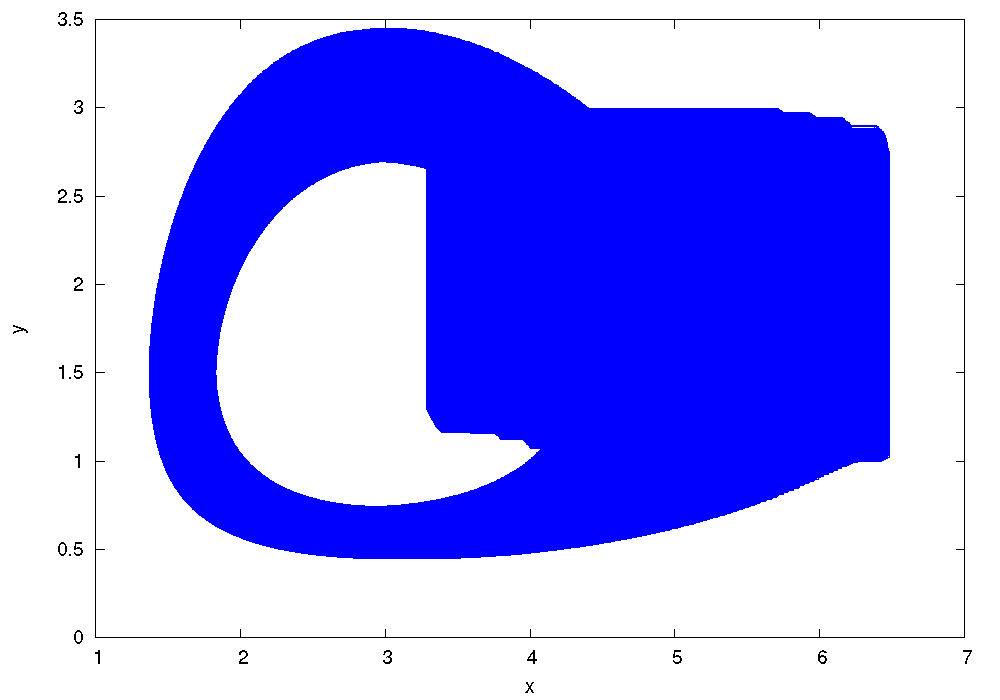}
		\caption{Flow*}
		\label{LotVolFlow}
	\end{subfigure}
	\begin{subfigure}[b]{0.21\linewidth}
			\centering
			\includegraphics[height=18mm,keepaspectratio]{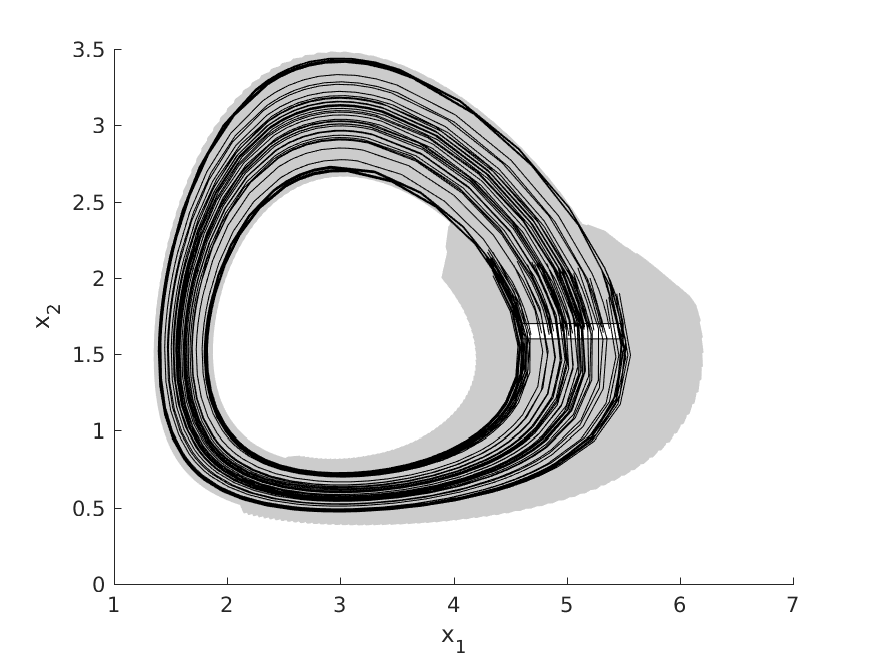}
			\caption{CORA}
			\label{VanderPolCORA}
	\end{subfigure}
	\begin{subfigure}[b]{0.21\linewidth}
	\centering
	\includegraphics[height=18mm,keepaspectratio]{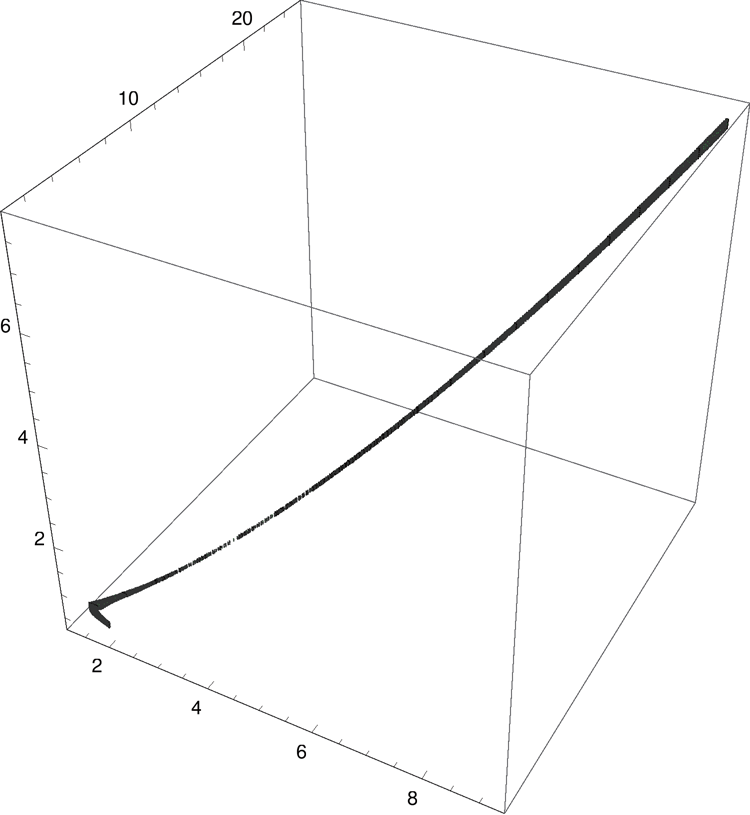}
	\caption{PRBT}
	\label{controller3d}
\end{subfigure}
	\caption{Lotka-Volterra: (a), (b), (c); controller 3D: (d)}
	\label{Lotka-Volterra}
\end{figure}

\begin{figure}[t!]
	\begin{subfigure}[b]{0.23\linewidth}
		\centering
		\includegraphics[height=18mm,keepaspectratio]{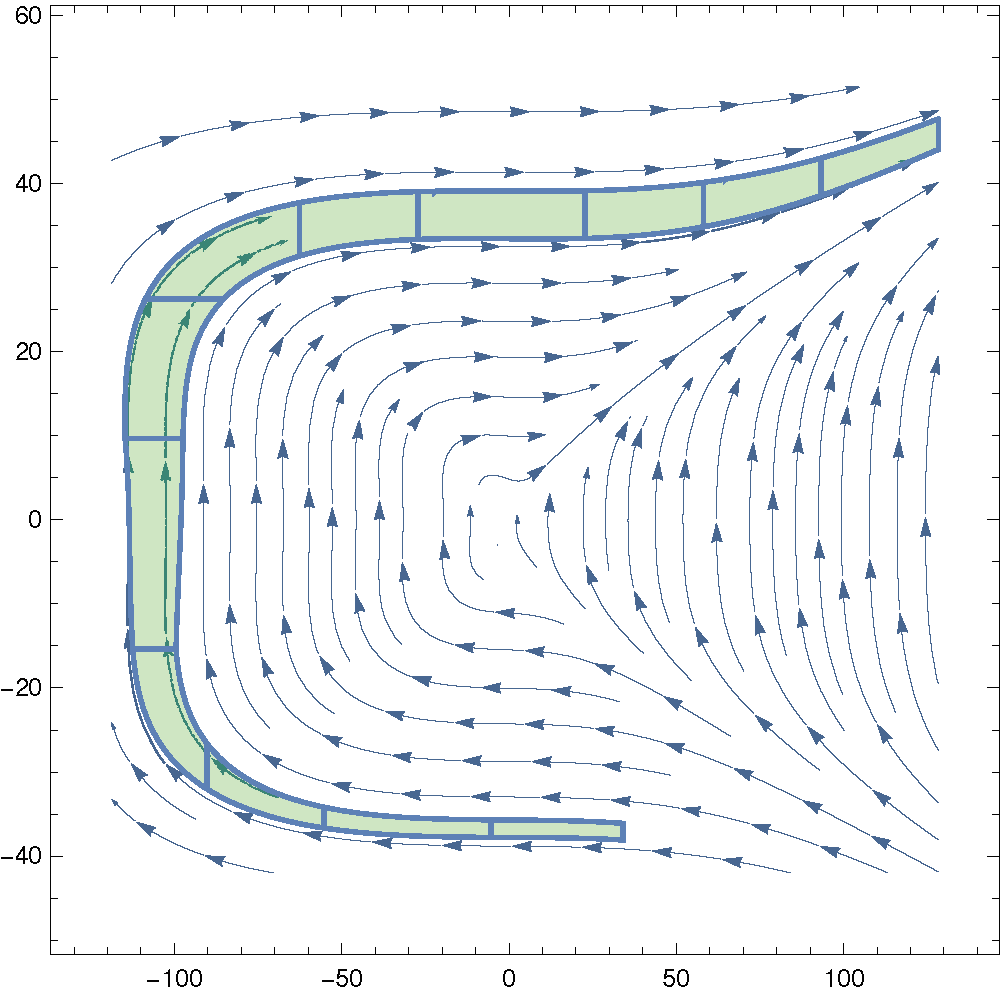}
		\caption{PRBT}
		\label{Controller2D}
	\end{subfigure}
	\begin{subfigure}[b]{0.23\linewidth}
		\centering
		\includegraphics[height=18mm,keepaspectratio]{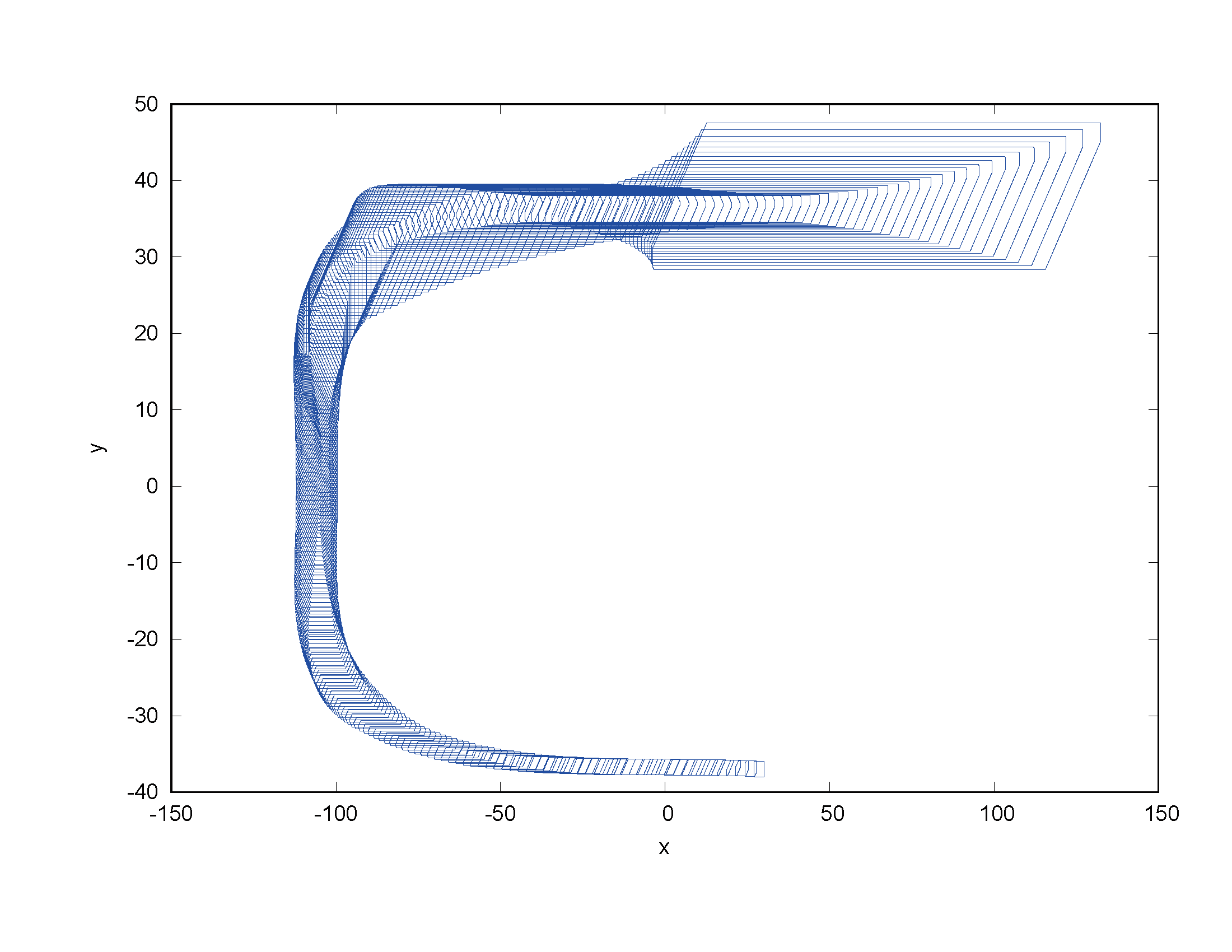}
		\caption{Flow*}
		\label{VandePolFlowlCORA}
	\end{subfigure}
	\begin{subfigure}[b]{0.23\linewidth}
	\centering
	\includegraphics[height=18mm,keepaspectratio]{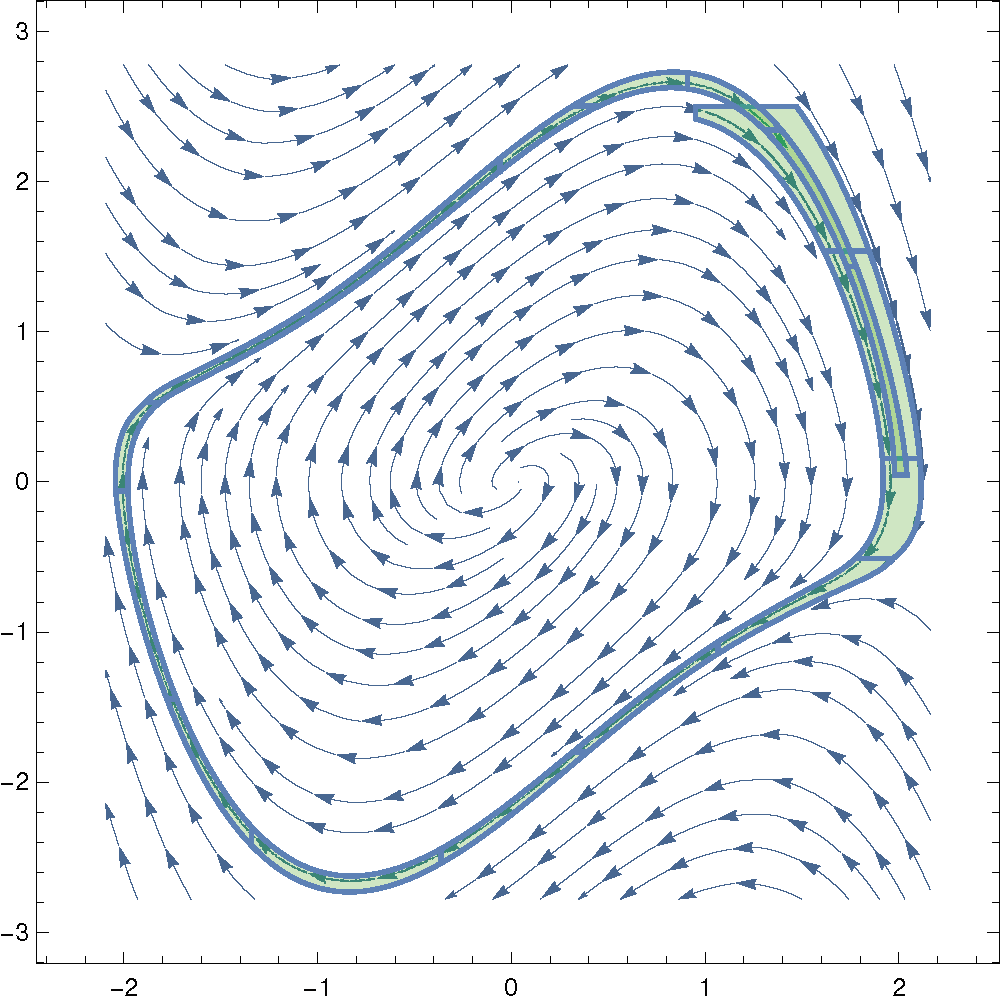}
	\caption{PRBT}
	\label{vanderpolPRBT}
	\end{subfigure}
	\begin{subfigure}[b]{0.23\linewidth}
	\centering
	\includegraphics[height=18mm,keepaspectratio]{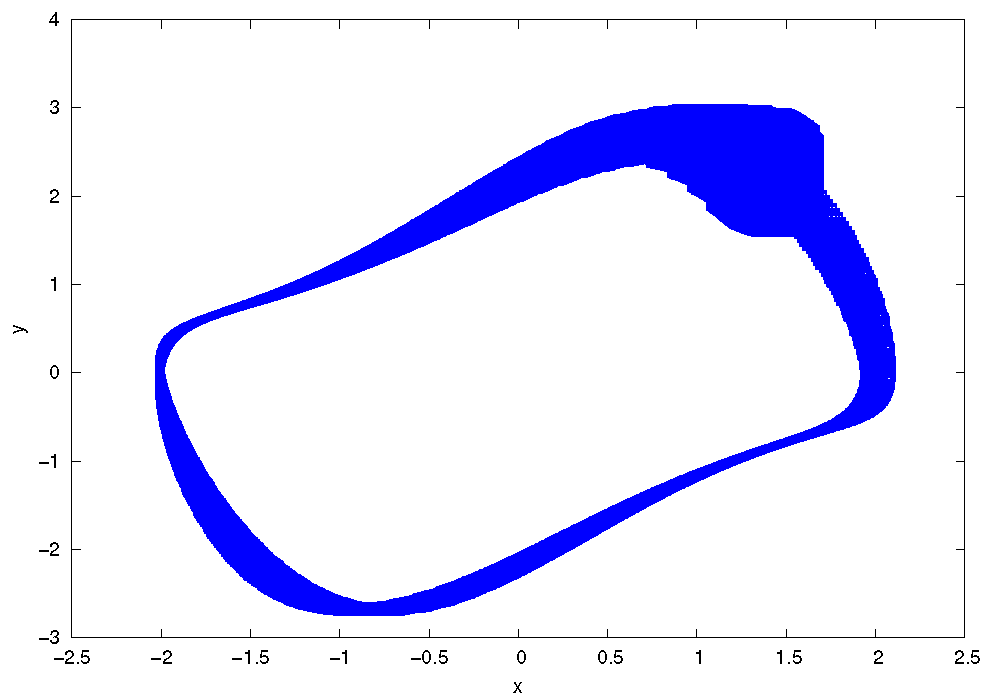}
	\caption{Flow*}
	\label{VandePolFlow}
	\end{subfigure}

	\caption{Controller 2d: (a), (b); Van der Pol Oscillator: (c), (d)} 
    \label{Flowgroup1}
\end{figure}

\begin{figure}[t!]
	\begin{subfigure}[b]{0.23\linewidth}
		\centering
		\includegraphics[height=20mm,keepaspectratio]{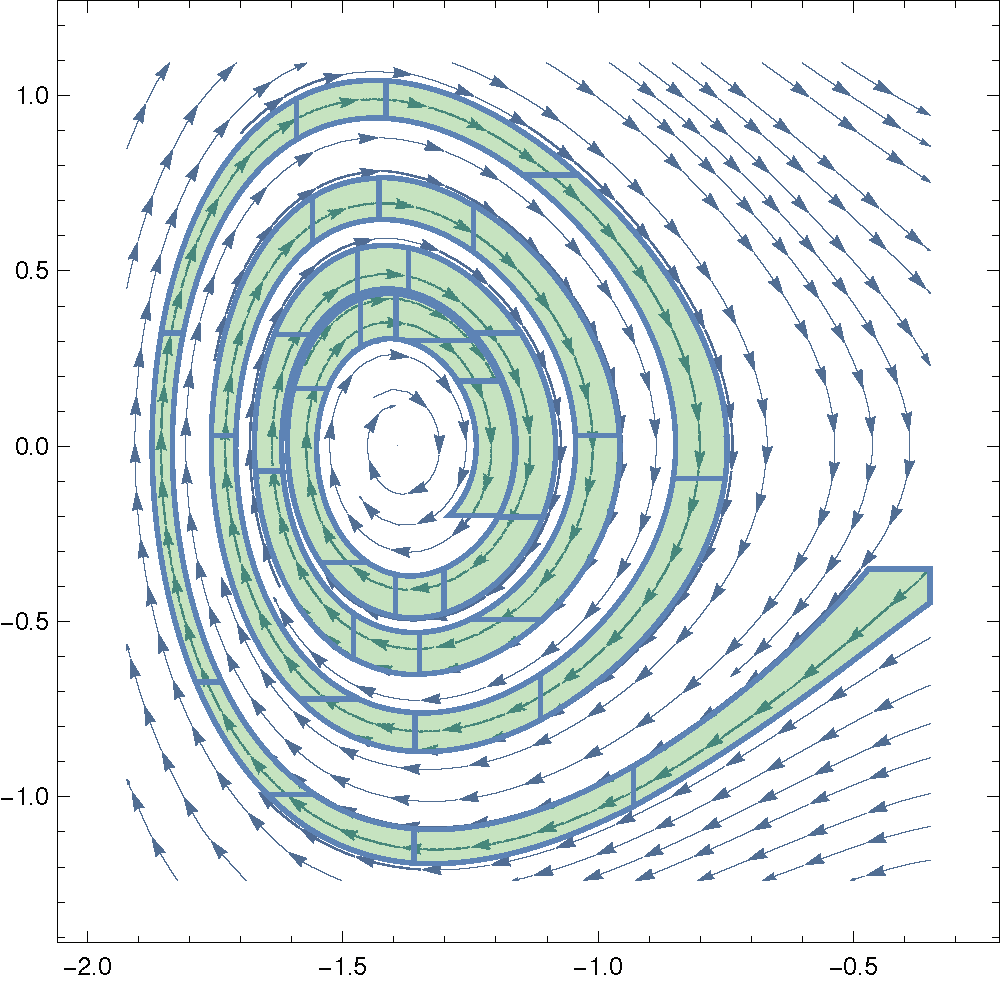}
		\caption{PRBT}
		\label{bucklingPRBT}
	\end{subfigure}
	\begin{subfigure}[b]{0.23\linewidth}
		\centering
		\includegraphics[height=20mm,keepaspectratio]{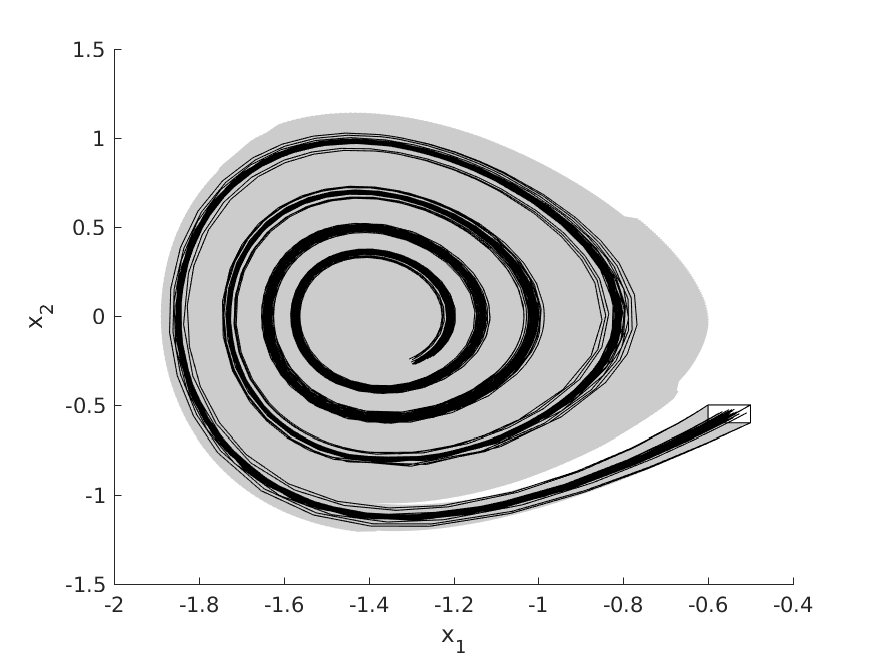}
		\caption{CORA}
		\label{bucklingpertCORA}
	\end{subfigure}
	\begin{subfigure}[b]{0.23\linewidth}
		\centering
		\includegraphics[height=20mm,keepaspectratio]{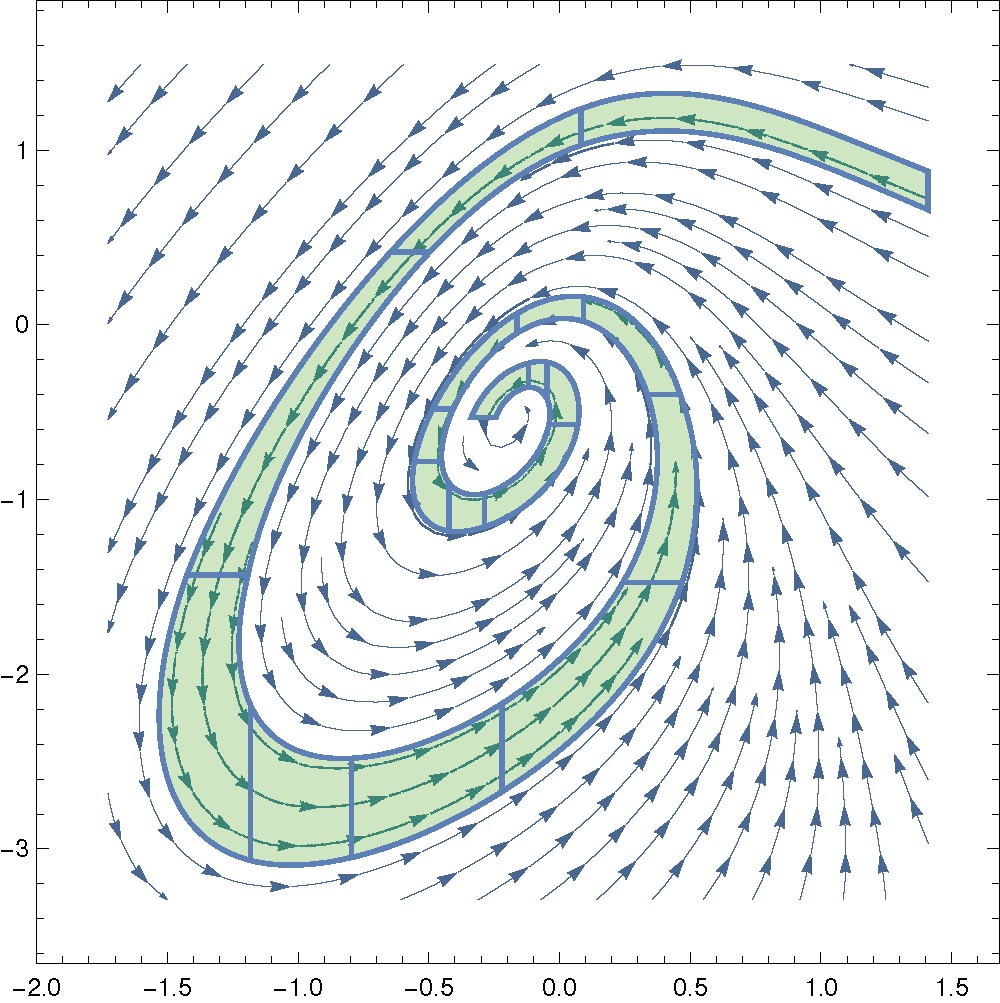}
		\caption{PRBT}
		\label{JetEnginePRBT}
	\end{subfigure}
	\begin{subfigure}[b]{0.23\linewidth}
		\centering
		\includegraphics[height=20mm,keepaspectratio]{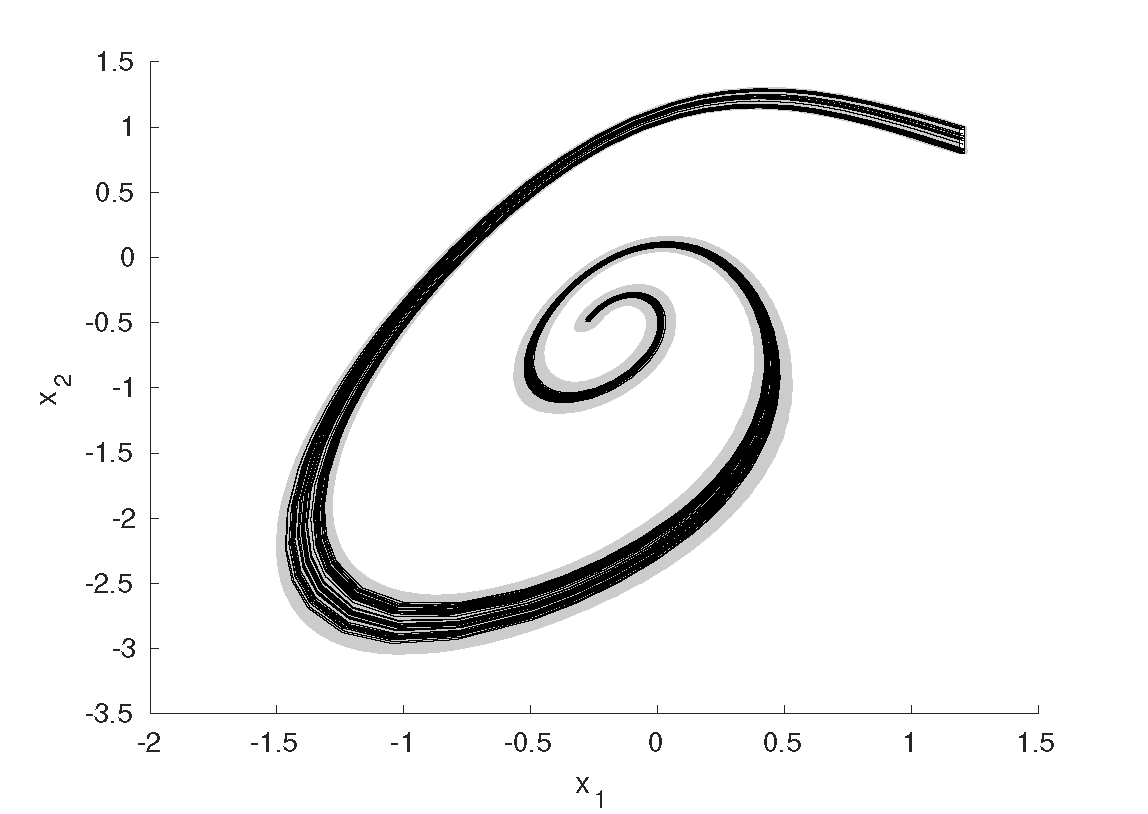}
		\caption{CORA}
		\label{JetEngineCORA}
	\end{subfigure}
\caption{Buckling Column: (a), (b); Jet Engine: (c), (d)}
\label{flowgroup2}
\end{figure}

\begin{remark}
Table~\ref{tbl:data} shows us how brutal the reality of reachability analysis of nonlinear systems is and this gets even worse in the presence of uncertainty and large initial set. As can be seen in Table~\ref{tbl:data}, both Flow* and CORA failed to give a solution for half of the benchmarks either due to timeout or due to exception. This phenomenon may be alleviated if smaller initial sets are provided or uncertainty is removed. In terms of computing time $T$, PRBT does not always outperform the other two tools. Actually, Flow* or CORA can be much faster in some cases. However, when the box transformation time for Taylor model was taken into account, the total computing time $TT$ of Flow* increased significantly. One point to note here is that, PRBT, in general, produces a much smaller number $N$ of flowpipe segments than the other two, which means that the time used to check the intersection of flowpipe with the unsafe set can be reduced considerably. In addition, as shown in Figure~\ref{Lotka-Volterra}--\ref{flowgroup2}, PRBT is more precise than the other two on average.  
\end{remark}

%

\subsection{Nonlinear Hybrid System}\label{sec:hybrid}
We use the tunnel diode oscillator (TDO) circuit (with different setting) introduced in~\cite{hartong2002model} to illustrate the application of our approach to hybrid system. The two state space variables are the voltage $x_1 = V_C$ across capacitor and the current $x_2 = I_L$ through the inductor. The system dynamics is described as follows,
\begin{equation*}
\dot{x}_1 = \frac{1}{C}(-h(x_1) + x_2) \qquad \dot{x}_2 = \frac{1}{L}(-x_1 - \frac{x_2}{G} + V_{in})
\end{equation*}

 where $h(x_1)$ describes the tunnel diode characteristic and $V_{in} = 0.3V$, $G = 5m\ {ohm}^{-1}$, $L = 0.5\mu H$ and $C = 2pF$. 

For this model, we want to define an initial state region $\Init$ which can guarantee the oscillating behaviour for the system. Due to the highly nonlinear behaviour of the system, a common strategy to deal with this model is to use a hybridized model to approximate the dynamics system and then apply formal verification to the hybrid model\cite{frehse2005verification,gupta2003towards}. In our experiment, we use three cubic equations to approximate the curve of $h(x_1)$. 
 \begin{equation*}
 h(x_1) = 
 \left
 \{\begin{array}{lc} 0.000847012 + 35.2297 x_1 - 395.261 x_1^2 + 1372.29 x_1^3, \ 0 \leq x_1 \leq 0.0691 \\
 1.242 - 0.033 x_1 - 47.4311 x_1^2 + 116.48 x_1^3,\ 0.0691 \leq x_1 \leq 0.3 \\
 -16.544 + 139.64 x_1 - 389.245 x_1^2 + 359.948 x_1^3,\ 0.3 \leq x_1 \leq 0.50 \\
 \end{array}\right.
 \end{equation*}

\begin{figure}[!t]
	\centering
	\includegraphics[scale=0.30,trim=0cm 20cm 0cm 2cm,clip]{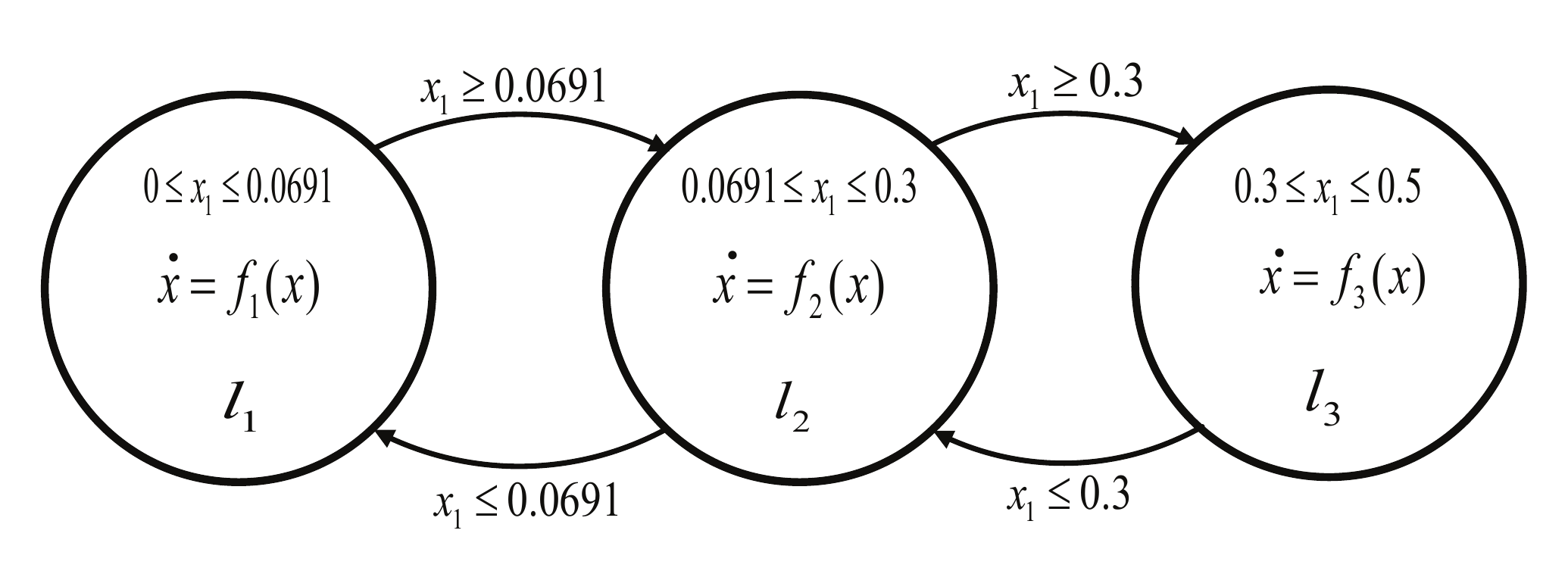}
	\caption{Hybridized model of TDO}
	\label{tunneldiodehybrid}
\end{figure}

From the piecewise function $h(x_1)$, we can derive a $3$-mode hybrid system which is shown in Figure~\ref{tunneldiodehybrid}. The system switches between the locations as the value of $x_1$ changes. 

\begin{figure}[t!]
	\begin{subfigure}[b]{0.3\linewidth}
		\centering
		\includegraphics[height=20mm,keepaspectratio]{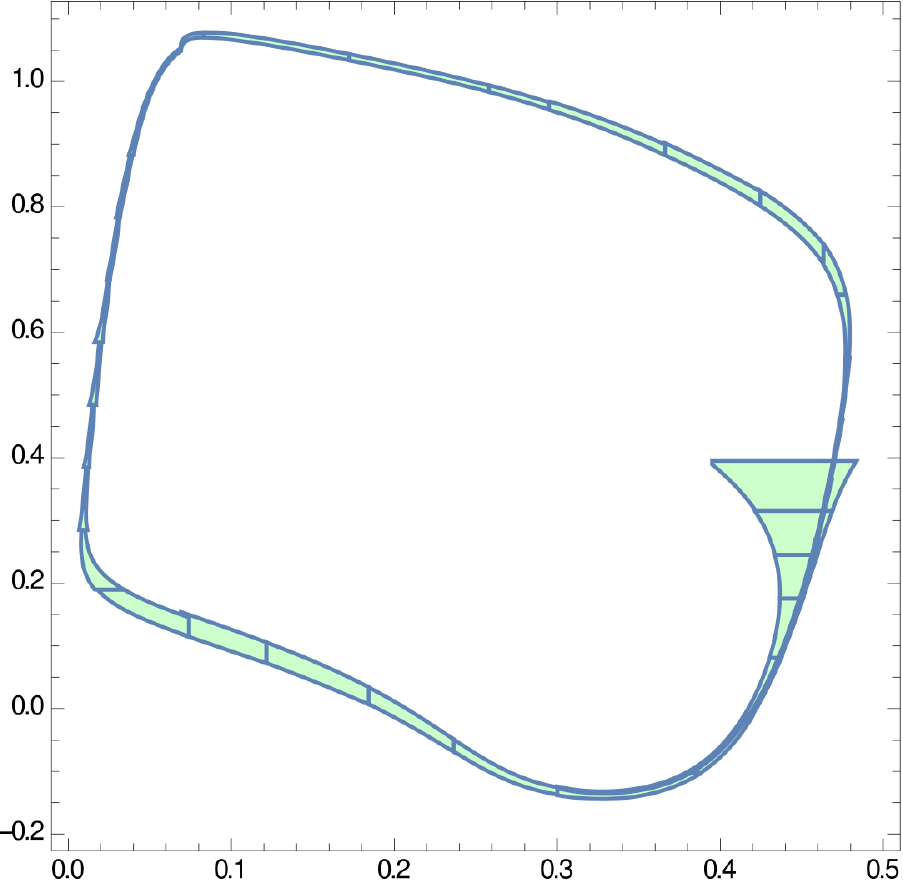}
		\caption{PRBT}
		\label{LABELPRBTSOLVER}
	\end{subfigure}
	\begin{subfigure}[b]{0.3\linewidth}
		\centering
		\includegraphics[height=20mm,keepaspectratio]{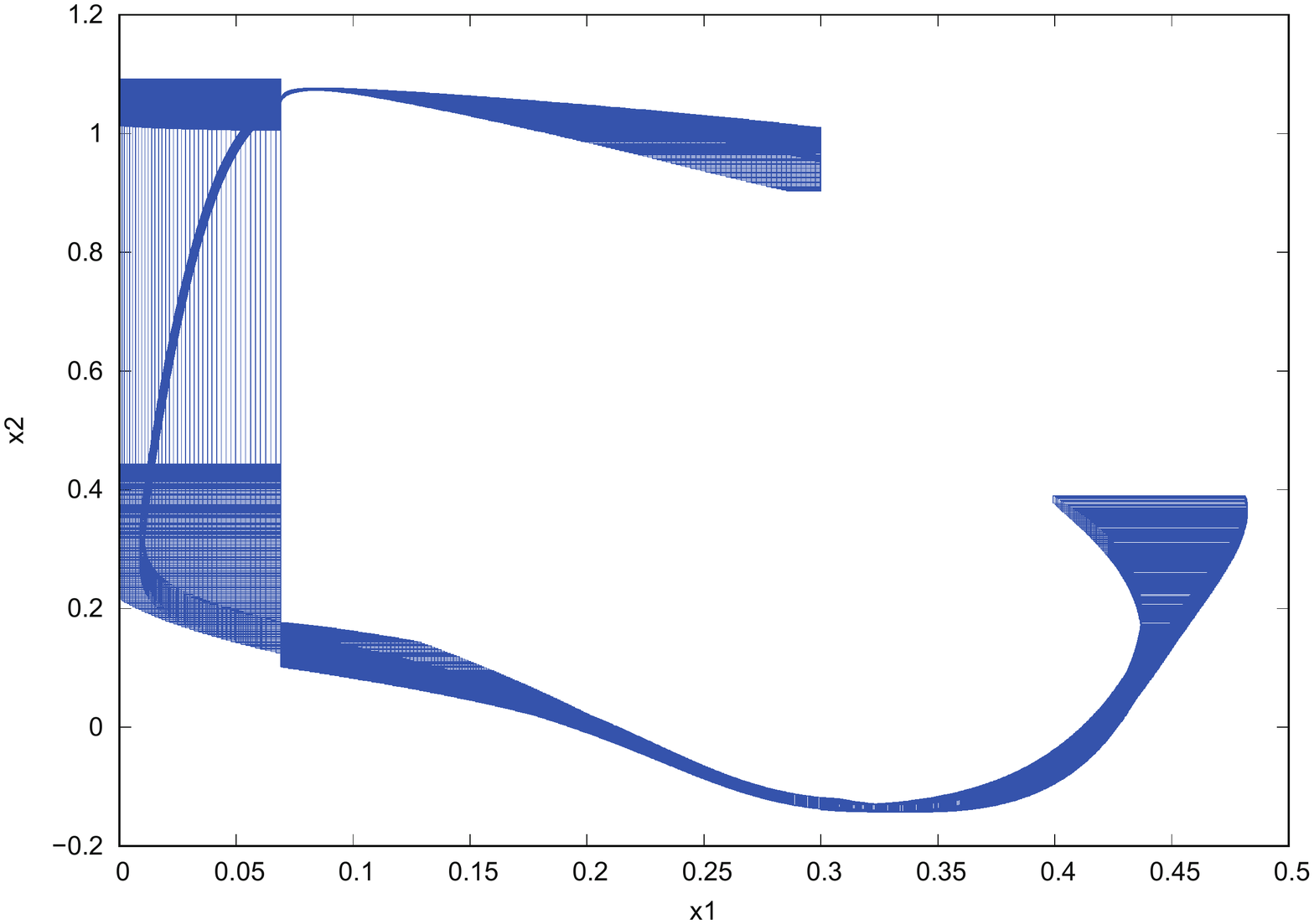}
		\caption{Flow*}
		\label{LABELFLOWSTAR}
	\end{subfigure}
    \begin{subfigure}[b]{0.3\linewidth}
	\centering
	\includegraphics[height=20mm,keepaspectratio]{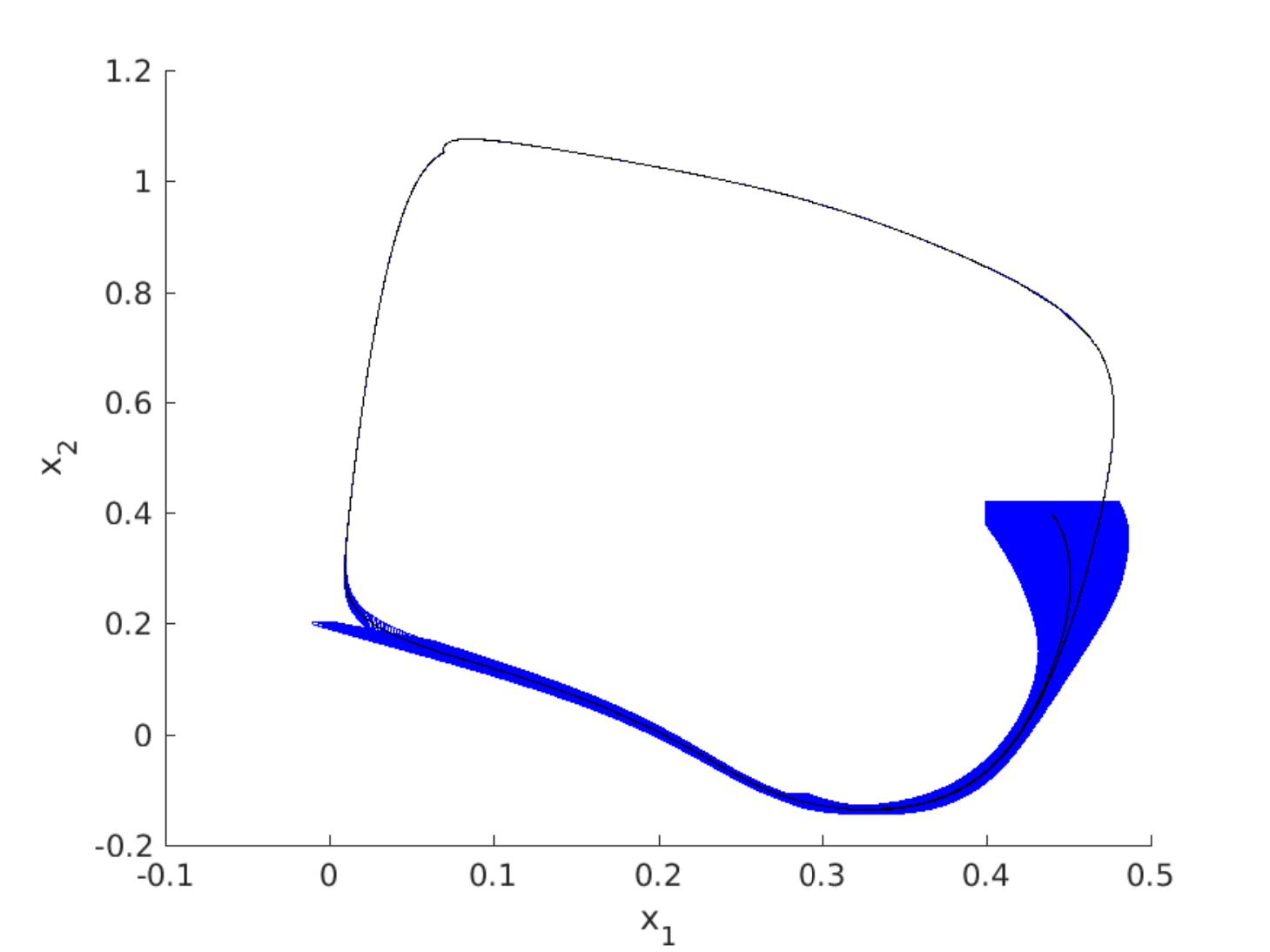}
	\caption{CORA}
	\label{LABELCORA}
\end{subfigure}
	\caption{Flowpipe of hybridized TDO}
	\label{hybridflow}
\end{figure}

Let the initial set be $\Init = \{ (x_1,x_2)\in\R^2 \mid 0.40 \leq x_1 \leq 0.48, 0.38 \leq x_2 \leq 0.39 \}$ on location $l_3$, we compute an overapproximation for the flowpipe using PRBT, Flow* and CORA respectively. As illustrated in Figure~\ref{hybridflow}, both PRBT and CORA found an invariant with roughly the same precision, which indicates the model oscillates for the initial set, while Flow* ran into an error. 

\vspace{-2ex}
\section{Conclusion}\label{sec:conclusion}
We propose a novel method to compute efficiently 
a tight overapproximation (flowpipe) of the reachable 
set for nonlinear continuous and hybrid systems with uncertainty. 
Our approach avoid the use of interval 
method for enclosure-box computation, which can produce bigger enclosure-boxes 
and fewer flowpipe segments. 
Experiments on several benchmark systems show 
that our method is more efficient and precise than other popular 
approaches proposed in literature.

%
%


\bibliographystyle{plain}
\bibliography{bibtex}
  
  
\end{document}